\documentstyle[epsf,psfig]{l-aa}
\def\erg{$erg ~cm^{-2} ~s^{-1}$ }
\def\deg{ ^\circ }
\def\arcm{$^{\prime}\,$}
\def\arcs{$^{\prime\prime}\,$}
\def\errx{$\varepsilon_{\rm{x}}$}
\begin{document}
\thesaurus{ 11(11.04.1; 11.12.2) }
\title{The ROSAT deep survey
\thanks{based on observations collected at the European Southern
Observatory, La Silla, Chile.}} 
\subtitle{ V. X-ray Sources and Optical Identifications in the Marano Field}
%
%
\author{
G.Zamorani\inst{1,2}
\and
M.Mignoli\inst{1}
\and
G. Hasinger\inst{3}
\and
R.Burg\inst{4}
\and
R.Giacconi\inst{5}
\and
M.Schmidt\inst{6}
\and
J.Tr\"umper\inst{7}
\and
P.Ciliegi\inst{1}
\and
C.Gruppioni\inst{1}
\and
B.Marano\inst{1,8}
}
%
%
\institute{ 
Osservatorio Astronomico di Bologna, 
Via Ranzani 1, 40127 Bologna, Italy
\and
Istituto di Radioastronomia del CNR, 
via Gobetti 101, 40129 Bologna, Italy
\and
Astrophysikalisches Institut Potsdam, An der Sternwarte 16, D-14482 Potsdam,
Germany
\and
Johns Hopkins University, Baltimore, MD 21218, USA
\and
European Southern Observatory, Karl-Schwarzschild-Str. 1,D-85748 Garching
bei M\"unchen, Germany
\and
California Institute of Technology, Pasadena, CA 91125,USA
\and
Max-Planck-Institut f\"ur extraterrestrische Physik, Karl-Schwarzschild-Str. 2,
D-85748 Garching bei M\"unchen, Germany
\and
Dipartimento di Astronomia, Universit\`a di Bologna, Via Ranzani 1, 
40127 Bologna, Italy
}
%
%
\offprints{G. Zamorani (zamorani@bo.astro.it)}
\date{Received 21 January 1999 / Accepted 26 March 1999}
\maketitle
\markboth {G.Zamorani et al.:
X-ray Sources and Optical Identifications in the Marano Field}{}
%
%
\begin{abstract}

We present the X-ray data and the optical identifications for
a deep ROSAT PSPC observation in the ``Marano field''. 
In the inner region of the ROSAT field (15\arcm \ radius) we detected 
50 X-ray sources with $S_x \ge 3.7 \times 10^{-15}$ \erg. When corrected 
for the different sensitivity over the field, the estimated observed surface 
density at $S_x \ge 4 \times 10^{-15}$ \erg is $272 \pm 40$ sources/sq.deg. 
Four X-ray sources, corresponding to 8\% of the total sample, have been
detected in radio images with a flux limit of about 0.2 mJy.

Careful statistical analysis of multi--colour CCD data in the error boxes of 
the 50 X-ray sources has
led to the identification of 42 sources, corresponding to 84\% of the
X-ray sample.  These 42 reliable identifications are 33 AGNs (including 
two radio galaxies and one BL Lac candidate; 79\% of the identified sources), 
2 galaxies, 3 groups or clusters of galaxies and 4 stars. 
If we divide our sample into two equally populated sub--samples as a 
function of flux, at $S_x = 6.5 \times 10^{-15}$ \erg, we find that
the percentage of identifications remains approximately constant (88\% and
80\% in the high and low flux sub--samples, respectively). 
AGNs are the dominant class of objects in both sub--samples (90\% of the 
optical identifications in the high flux sub--sample and 65\% in the low flux
sub--sample), while the few identifications with clusters and galaxies are
all in the low flux sub--sample. 

We also show that it is likely that a few of
the 8 unidentified sources are such because the derived X-ray positions
may be offset with respect to the real ones due to confusion effects.
The unidentified sources have a large ratio of X-ray to optical fluxes and
most of them have harder than average X-ray spectra. Since
most of the identified objects with these characteristics in our field and
in the Lockman field are AGNs, we conclude that also most of these sources
are likely to be AGNs.

Finally, comparing the optically and X-ray selected samples of AGNs in this
field, we estimate that the ``efficiency'' of AGN selection with X-ray
exposures reaching about $4 \times 10^{-15}$ \erg is
$\sim$65\% and $\sim$20\% in the magnitude ranges $m_B < 22.5$ and 
$22.5 < m_B < 23.5$, respectively. On the other hand, a not negligible
fraction of the X-ray selected AGNs would have not been easily selected
as AGN candidates on the basis of purely optical criteria, either because
of colours similar to those of normal stars or because of morphological
classification not consistent with that of point--like sources.
\end{abstract}
%
%
\section{Introduction}

Complete or, at least, statistically well defined samples of optical
identifications of faint X-ray sources, are important for a number
of scientific goals. For example, combined optical and X-ray data allow to
obtain information about the luminosity functions of various types of
X-ray sources as well as their evolution with redshift. In turn, these
informations can be used to further constrain models for the production of the
X-ray background, discovered more than thirty years ago (Giacconi et al.
1962). There is a general consensus that the majority of the optically 
identified X-ray sources, at least for fluxes 
$ S_{0.5 - 2 keV} \ge 5 \times 10^{-15}$ \erg, are active galactic nuclei 
(AGNs), i.e. quasars and Seyfert galaxies with broad emission lines 
(Shanks et al. 1991, Georgantopoulos et al. 1996, McHardy et al. 1998, 
Hasinger et al. 1998 (Paper I) and Schmidt et al. 1998 (Paper II)).

In the first two papers of this series we have presented the 
X-ray and optical data
for a complete catalogue of 50 X-ray sources with PSPC fluxes (0.5-2 keV)
above $ 5.5 \times 10^{-15}$ \erg in the Lockman Field. Optical and X-ray
results at a much fainter flux limit in this field, which has been observed
with the deepest ROSAT PSPC and HRI observations, will be presented
elsewhere. 

In this paper we present and discuss the X-ray data (Section 2)
and the optical data (Section 3) for 50 sources detected with the PSPC at a 
flux limit $ S_x \ge 3.7 \times 10^{-15}$ \erg in the Marano Field. 
A discussion of the main results (percentages of identifications with
different classes of objects, hardness ratio as a function of X-ray flux,
comparison between X-ray and optically selected AGNs) is given in Section 4.
Throughout the paper we use $H_0$ = 50 km s$^{-1}$ Mpc$^{-1}$ and $q_0$ = 0.5.
\section{The X-ray Data}

\subsection{The sample}

The ROSAT--PSPC pointed observations of the Marano field, centered at
\hbox{RA = $ 03^h 15^m 09^s $}, \hbox{DEC = -55$\deg$ 13\arcm 57\arcs}
(epoch 2000.0), have been carried out in the time interval 
December~1992 -- July~1993, for a total of 56 ksec of 
observing time. The observations were performed in the ``wobble mode''. 

The X-ray analysis (e.g. source detection, position and flux determination 
of the detected sources, determination of sensitivity as a function of distance 
from the center, etc.) has been done applying a maximum likelihood
method. For details about the various steps in this analysis we refer the
reader to Hasinger et al. (1993 and 1998). Although detection of the
sources has been run in various ROSAT bands, we report here results
only for the hard band (0.5--2 keV), which has been shown (Hasinger
et al. 1993) to be the most efficient for point source detection.

\begin{table*}
\caption[]{ The X-ray Data}
\begin{tabular}{ccccrrccr} 
\hline\noalign{\smallskip}

{\#} & RA (2000) & DEC (2000) & Err & Off-axis & ML & Net Counts & S$_x$ & HR~~~~~~\\
     &           &           &[arcsec]&[arcmin]&    & &[$\times 10^{-14}$ cgs]&    \\
\noalign{\smallskip}
\hline\noalign{\smallskip}
  ~~X013--01 & 3 13 47.1 & $-$55 11 48 &  1.6 & 12.1~~ & 448.2 & 210.1 & 5.00 $\pm$ 0.42 &  $-$0.03 $\pm$ 0.06       \\
  ~~X012--02 & 3 13 29.2 & $-$55 10 20 &  2.7 & 14.8~~ & 203.3 & 155.7 & 3.82 $\pm$ 0.37 &     0.41 $\pm$ 0.12       \\
  ~~X046--03 & 3 14 32.4 & $-$55 14 43 &  1.6 &  5.5~~ & 303.1 & 130.1 & 2.99 $\pm$ 0.32 &     0.20 $\pm$ 0.10       \\
  ~~X036--04 & 3 16 50.3 & $-$55 11 10 &  2.4 & 14.5~~ & 175.4 & 118.9 & 2.91 $\pm$ 0.30 &  $-$0.10 $\pm$ 0.08       \\
  ~~X021--05 & 3 14 56.4 & $-$55 20 08 &  1.8 &  6.7~~ & 190.4 & 102.3 & 2.36 $\pm$ 0.25 &     0.88 $\pm$ 0.20       \\
  ~~X025--06 & 3 15 49.5 & $-$55 18 11 &  1.8 &  7.1~~ & 184.7 &  99.8 & 2.31 $\pm$ 0.30 &     0.28 $\pm$ 0.13       \\
  ~~X027--07 & 3 16 05.5 & $-$55 15 44 &  2.2 &  8.1~~ & 124.6 &  71.8 & 1.66 $\pm$ 0.28 &  $-$0.21 $\pm$ 0.10       \\
  ~~X041--08 & 3 15 28.7 & $-$55 10 32 &  2.6 &  4.1~~ &  99.7 &  62.0 & 1.42 $\pm$ 0.25 &     0.58 $\pm$ 0.15       \\
  ~~X240--09 & 3 16 38.2 & $-$55 06 41 &  3.9 & 14.4~~ &  56.1 &  50.9 & 1.25 $\pm$ 0.24 &     0.45 $\pm$ 0.25       \\
  ~~X033--10 & 3 15 58.3 & $-$55 26 40 &  4.2 & 14.6~~ &  46.7 &  50.9 & 1.25 $\pm$ 0.23 &  $-$0.48 $\pm$ 0.11       \\
  ~~X042--11 & 3 15 40.4 & $-$55 12 25 &  2.3 &  4.5~~ &  77.1 &  52.4 & 1.20 $\pm$ 0.18 &     0.32 $\pm$ 0.32       \\
  ~~X043--12 & 3 15 09.8 & $-$55 13 18 &  2.3 &  0.5~~ &  88.2 &  48.5 & 1.10 $\pm$ 0.25 &     0.37 $\pm$ 0.32       \\
  ~~X023--13 & 3 15 25.2 & $-$55 18 28 &  2.7 &  5.2~~ &  59.1 &  43.5 & 1.00 $\pm$ 0.17 &  $-$0.27 $\pm$ 0.19       \\
  ~~X108--14 & 3 15 37.9 & $-$55 01 42 &  4.4 & 12.7~~ &  30.8 &  41.0 & 0.98 $\pm$ 0.20 &     0.63 $\pm$ 0.62       \\
  ~~X030--15 & 3 16 26.1 & $-$55 23 00 &  4.3 & 14.2~~ &  28.3 &  39.6 & 0.97 $\pm$ 0.18 &     0.03 $\pm$ 0.20       \\
  ~~X304--16 & 3 15 11.4 & $-$55 09 30 &  3.1 &  4.3~~ &  56.5 &  40.8 & 0.94 $\pm$ 0.18 &     0.08 $\pm$ 0.27       \\
  ~~X019--17 & 3 14 21.8 & $-$55 23 55 &  3.6 & 12.3~~ &  43.0 &  38.8 & 0.93 $\pm$ 0.27 &     0.24 $\pm$ 0.31       \\
  ~~X029--18 & 3 16 29.9 & $-$55 19 06 &  3.8 & 12.5~~ &  30.1 &  37.0 & 0.89 $\pm$ 0.17 &     0.14 $\pm$ 0.30       \\
  ~~X039--19 & 3 15 52.9 & $-$55 08 20 &  3.2 &  8.1~~ &  47.5 &  38.0 & 0.88 $\pm$ 0.18 &     1.00~~~~~~~~~~	     \\
  ~~X001--20 & 3 15 20.7 & $-$55 02 33 &  3.6 & 11.3~~ &  46.4 &  37.4 & 0.88 $\pm$ 0.26 &  $-$0.07 $\pm$ 0.24       \\
  ~~X049--21 & 3 15 06.0 & $-$55 09 42 &  3.0 &  4.1~~ &  49.4 &  35.9 & 0.82 $\pm$ 0.21 &  $-$0.19 $\pm$ 0.20       \\
  ~~X211--22 & 3 13 45.5 & $-$55 19 25 &  4.4 & 13.4~~ &  21.5 &  33.8 & 0.81 $\pm$ 0.19 &  $-$0.08 $\pm$ 0.32       \\
  ~~X404--23 & 3 16 48.9 & $-$55 12 40 &  9.3 & 14.1~~ &  10.2 &  29.7 & 0.73 $\pm$ 0.21 &     1.00~~~~~~~~~~	     \\
  ~~X031--24 & 3 15 48.7 & $-$55 22 50 &  4.4 & 10.6~~ &  17.4 &  30.2 & 0.71 $\pm$ 0.18 &  $-$0.21 $\pm$ 0.29       \\
  ~~X050--25 & 3 15 07.6 & $-$55 04 58 &  3.8 &  8.8~~ &  10.0 &  30.1 & 0.70 $\pm$ 0.16 &  $-$0.26 $\pm$ 0.23       \\
  ~~X235--26 & 3 16 31.7 & $-$55 12 27 &  5.2 & 11.7~~ &  14.3 &  26.7 & 0.63 $\pm$ 0.18 &     0.21 $\pm$ 0.45       \\
  ~~X045--27 & 3 15 10.7 & $-$55 15 23 &  3.5 &  1.6~~ &  30.8 &  26.9 & 0.61 $\pm$ 0.15 &     0.24 $\pm$ 0.33       \\
  ~~X409--28 & 3 14 26.3 & $-$55 17 41 &  5.2 &  7.4~~ &  22.0 &  26.2 & 0.61 $\pm$ 0.14 &     0.84 $\pm$ 0.85       \\
  ~~X301--29 & 3 14 36.3 & $-$55 14 04 &  4.1 &  4.9~~ &  20.2 &  26.0 & 0.60 $\pm$ 0.15 &  $-$0.21 $\pm$ 0.27       \\
  ~~X408--30 & 3 14 50.3 & $-$55 19 39 &  4.1 &  6.6~~ &  17.1 &  25.6 & 0.59 $\pm$ 0.16 &     0.65 $\pm$ 0.87       \\
  ~~X207--31 & 3 13 50.3 & $-$55 13 00 &  4.9 & 11.5~~ &  16.1 &  24.7 & 0.59 $\pm$ 0.15 &  $-$0.12 $\pm$ 0.42       \\
  ~~X040--32 & 3 15 43.1 & $-$55 07 46 &  3.5 &  7.6~~ &  27.5 &  24.9 & 0.58 $\pm$ 0.14 &     0.34 $\pm$ 0.41       \\
  ~~X028--33 & 3 16 21.8 & $-$55 18 00 &  4.8 & 11.0~~ &  16.5 &  23.8 & 0.56 $\pm$ 0.14 &     0.24 $\pm$ 0.45       \\
  ~~X250--34 & 3 15 23.3 & $-$55 04 03 &  4.4 &  9.9~~ &  17.4 &  24.0 & 0.56 $\pm$ 0.14 &     0.65 $\pm$ 0.31       \\
  ~~X011--35 & 3 13 39.9 & $-$55 07 21 &  6.1 & 14.4~~ &  11.2 &  22.7 & 0.56 $\pm$ 0.15 &     0.04 $\pm$ 0.37       \\
  ~~X032--36 & 3 15 38.7 & $-$55 22 33 &  5.4 &  9.7~~ &  13.4 &  23.2 & 0.54 $\pm$ 0.14 &  $-$0.01 $\pm$ 0.43       \\
  ~~X251--37 & 3 15 31.0 & $-$55 04 43 &  6.0 &  9.5~~ &  16.8 &  22.5 & 0.52 $\pm$ 0.14 &     0.07 $\pm$ 0.67       \\
  ~~X024--38 & 3 15 34.7 & $-$55 19 27 &  5.4 &  6.7~~ &  12.4 &  22.7 & 0.52 $\pm$ 0.16 &  $-$0.11 $\pm$ 0.23       \\
  ~~X015--39 & 3 13 51.6 & $-$55 18 33 &  6.9 & 12.2~~ &  11.8 &  21.4 & 0.51 $\pm$ 0.14 &  $-$0.43 $\pm$ 0.22       \\
  ~~X236--40 & 3 16 24.0 & $-$55 11 44 &  4.3 & 10.7~~ &  15.9 &  21.3 & 0.50 $\pm$ 0.13 &     0.02 $\pm$ 0.32       \\
  ~~X234--41 & 3 16 23.7 & $-$55 15 17 &  5.3 & 10.6~~ &  10.7 &  20.7 & 0.48 $\pm$ 0.15 &  $-$0.15 $\pm$ 0.41       \\
  ~~X306--42 & 3 15 50.1 & $-$55 09 15 &  4.5 &  7.2~~ &  17.0 &  20.1 & 0.47 $\pm$ 0.14 &  $-$0.22 $\pm$ 0.38       \\
  ~~X051--43 & 3 15 01.6 & $-$55 03 40 &  6.6 & 10.2~~ &  11.1 &  19.9 & 0.46 $\pm$ 0.14 &     0.93 $\pm$ 1.00       \\
  ~~X407--44 & 3 14 12.4 & $-$55 25 52 &  7.3 & 14.7~~ &   9.8 &  18.2 & 0.45 $\pm$ 0.14 &  $-$0.33 $\pm$ 0.25       \\
  ~~X233--45 & 3 16 18.9 & $-$55 14 29 &  7.2 &  9.8~~ &   9.8 &  19.3 & 0.45 $\pm$ 0.14 &     0.26 $\pm$ 0.41       \\
  ~~X109--46 & 3 16 08.1 & $-$55 17 25 &  4.8 &  9.0~~ &  11.9 &  18.3 & 0.43 $\pm$ 0.12 &     0.05 $\pm$ 0.30       \\
  ~~X213--47 & 3 14 12.1 & $-$55 18 24 &  5.4 &  9.5~~ &  11.7 &  17.9 & 0.42 $\pm$ 0.14 &     1.00~~~~~~~~~~	     \\
  ~~X022--48 & 3 15 03.4 & $-$55 19 06 &  4.4 &  5.4~~ &  12.3 &  18.3 & 0.42 $\pm$ 0.12 &     1.00~~~~~~~~~~	     \\
  ~~X215--49 & 3 14 29.6 & $-$55 16 44 &  5.3 &  6.5~~ &  12.3 &  17.7 & 0.41 $\pm$ 0.12 &     1.00~~~~~~~~~~	     \\
  ~~X264--50 & 3 14 49.1 & $-$55 22 24 &  5.2 &  9.2~~ &  12.0 &  15.8 & 0.37 $\pm$ 0.14 &     1.00~~~~~~~~~~	     \\
\noalign{\smallskip}\hline
\end{tabular}
\end{table*}

Table 1 contains the X-ray data for the complete sample of sources
in a circular area with radius of 15\arcm. Within this area we detected 50
sources with a maximum likelihood value ML $\ge$ 9.8. With this 
adopted threshold in ML we expect less than one spurious source
over the entire area. 
The first column gives an identification code for the X-ray sources: the
first number identifies the sources in our working lists, while the second
one (from 1 to 50) represents the source rank in order of
decreasing X-ray flux. Columns 2 to 5 give
the $\alpha$ (2000) and $\delta$ (2000) positions, 
with the one sigma error on each coordinate (in arcsec) derived from the
maximum likelihood fitting, and the off--axis
distance from the center of the X-ray field (in arcmin). The next three 
columns give the maximum likelihood value in the hard band, the net counts and 
the corresponding 0.5 -- 2.0 keV flux and error.
The flux has been computed assuming the same intrinsic spectrum for all the
sources, i.e. a power law spectrum with a slope $\alpha$ = 1.0, with
no intrinsic absorption and a galactic absorption corresponding to
$N_H = 2.5 \times 10^{20}$. These assumed parameters correspond to the 
conversion factor  1 ct/s = 2.26 $\times 10^{-12} $ \erg at the center of
the field. The last column gives the hardness ratio with its statistical error.

Since most of the detected X-ray sources are too faint to obtain a full 
resolution spectrum, we have characterized their spectra using the hardness 
ratio technique. The hardness ratio is defined as HR=(H-S)/(H+S), 
where S~and~H~are the net counts in the PSPC energy channels 11--41 and
52--201, respectively, corresponding approximately to the energy ranges
0.1--0.4 and 0.5--2.0~keV. The source counts in
each band have been obtained summing all the counts from a circle centered at
the source position. The radius of the extraction circle was 60\arcs. However,
when the extraction circles of two sources overlapped, this radius was reduced
to 30\arcs in order to avoid contamination from the nearby source. 
The resulting net counts 
for the instrumental radial vignetting and for the energy dependence
of the ROSAT/PSPC Point Spread Function (Hasinger et al. 1994). The 
background counts were estimated in a circular area near the source position,
after excluding the contribution from near--by sources.
Five sources, corresponding to 10\% of the total sample, have a formal
hardness ratio equal to or greater 1.00 (i.e. zero or negative net counts
detected in the soft band). Four of them are at the faint flux limit of the
entire sample.

To eliminate a possible systematic error of a few arcsec in the X-ray
positions, we have cross--correlated the positions of the 50 X-ray sources
within 15\arcm \ from the center with those of 29 previously known optically 
selected AGNs with m$_B \le$ 22.5 in the same area (Zitelli et al. 1992; 
Mignoli et al. in preparation). We have found 19 positional coincidences, 
with distances smaller than 12\arcs between X-ray and optical positions, while 
less than 0.3 random coincidences are expected on a statistical basis. 
The average offset between X-ray and optical positions is about 4\arcs in right
ascension and less than 1\arcs in declination, with no additional trend as a 
function of position in the field. The one sigma uncertainty for the offset
in each coordinate is of the order of 0.9\arcs. 
We have therefore applied this average offset to all the original X-ray
positions and corrected the ML positional errors by adding in quadrature
the uncertainty on the offset. Both the positions and the positional errors
given in Table 1 are the ``corrected'' values. 

\begin{figure}
\vspace{0.5cm}
\vspace{8.7cm}
\caption[]{Gray scale representation of the hard ROSAT image. The image
has been slightly smoothed with a gaussian filter with $\sigma$ = 10\arcs. 
All the 50 sources detected within a radius of 15\arcm~are labelled.}
\end{figure}

Figure 1 shows a gray scale representation of the hard ROSAT image. The image
has been slightly smoothed with a gaussian with $\sigma$ = 10\arcs. All the 
sources detected within a radius of 15\arcm \ are labelled. The figure shows the
capability of the maximum likelihood algorithm in retrieving pairs of sources
relatively close to each other (see, for example, the two pairs 49--304 and
46--301, for which the distances between the X-ray positions are of the
order~of~50~arcsec). For distances between two X-ray centroids
smaller than $\sim$40\arcs, however, our detection algorithm
is unable to separate the two X-ray sources and will find a single
source (Hasinger et al. 1993), with a position intermediate between the two
true positions. In these cases the resulting X-ray position and its associated
error would not be reliable and this would lead to problems in the optical
identification process. 

In order to estimate how many such cases we may have in our list of sources 
we have simulated 1000 random
samples with the same number of detected sources and with the same radial
surface densities of sources as the real sample (i.e. higher density in the
inner region of the field and lower density in the outer region). For each
sample we have then computed the number of sources which have a nearby companion
at a distance smaller than the smallest observed distance in the real
sample (i.e. 48\arcs). Figure 2 shows the normalized histogram of the 
number of such sources in the 1000 random samples. The figure shows that in only
2\% of the cases there is no pair of sources with a distance smaller than
48~arcsec, while in the central 75\% of the cases (hatched area) the number of
such sources ranges from 4 to 10. 
\begin{figure}[t]
\epsfysize=9cm
\epsfbox{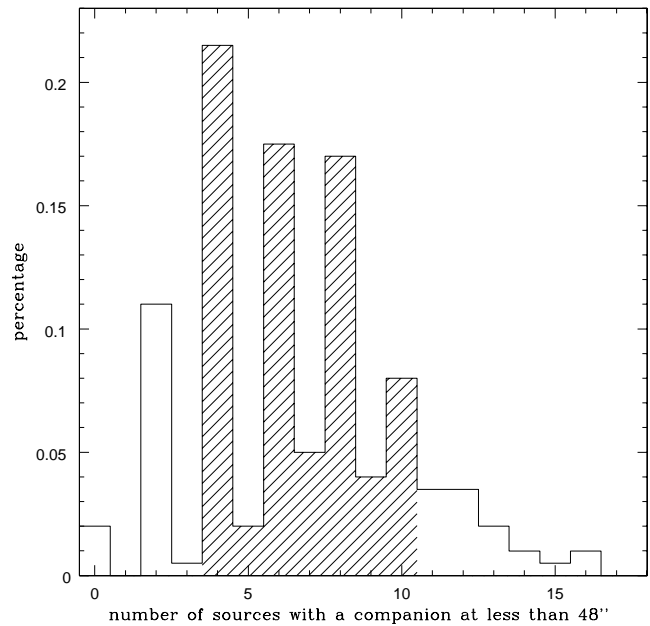}
\caption[]{Histogram of the number of sources with a companion at less than
48\arcs in 1000 simulated samples. In 75\% of the cases the number of these
sources is between 4 and 10 (hatched area).}
\end{figure}

Since each pair is counted twice in this histogram, this means that we may 
expect that in about 2--5 cases a 
single source in our list may be produced by two different close--by sources, so
that its position may be significantly wrong and we may not be able to find any
reliable optical counterpart within the error box.
Note that this estimate is based on purely geometric considerations; moreover,
since in computing the expected number of close--by sources 
we have adopted the observed surface density of sources, we are not 
considering in this order of magnitude estimate the cases in which one or both
the sources in a close pair are below the detection limit. 
Results from more detailed simulations, which take into account all
the possible effects of source confusion,  are discussed in Hasinger et al.
(1998).
\begin{figure}
\epsfysize=9cm
\epsfbox{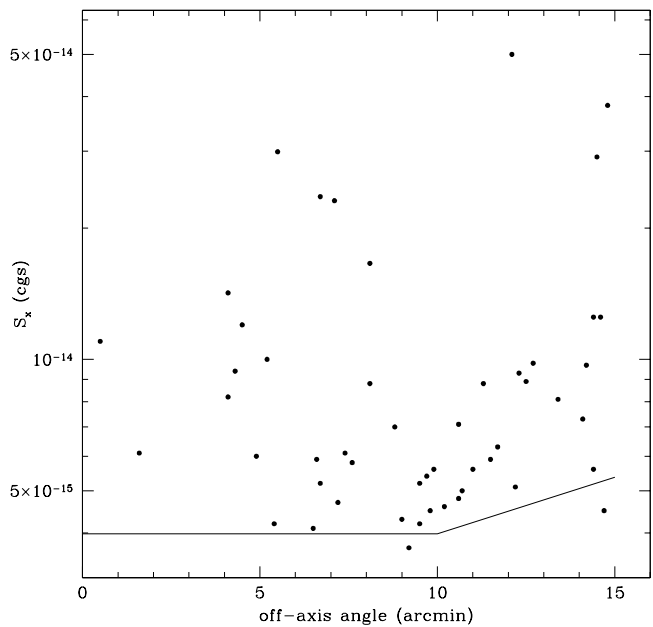}
\caption[]{Hard X-ray flux versus the off--axis angle for all
the X-ray sources listed in Table 1. The curve shows the 
limiting flux as a function of off--axis angle which has been used to
estimate the ``corrected'' observed surface density of sources.
}
\end{figure}

Figure 3 shows the hard X-ray flux versus the off--axis angle for all
the X-ray sources listed in Table 1. Even if we have considered only sources
within 15\arcm \ from the center, the limiting flux
is not constant over the adopted field of view. Its increase at distances
greater than about 10\arcm \ is mainly due to the
increase of the width of the point spread function with the off-axis angle.
The curve drawn in the figure shows our estimated limiting flux as a function 
of the off--axis angle for the complete sample. 
When corrected for the different sensitivity over the field, 
and using only the 48 sources with flux greater than the adopted limiting
flux, the estimated observed surface density at 
$S_x \ge 4 \times 10^{-15}$ \erg is $272 \pm 40$ sources/sq.deg.
This value is consistent with the surface density
(248 sources/sq.deg.) 
derived by Hasinger et al. (1993) from the composite observed log~N  -- log~S 
relationship using the deeper ROSAT data in the Lockman Hole together
with a number of shallower fields. 

\section{The Identification of the X-ray Sources}
\subsection{The Radio Data}

We observed the Marano field at 1.4 and 2.4 GHz
with the Australia Telescope Compact Array (ATCA). 
The radio observations were carried out on 1994 January 4, 5, 6 and 7.
Details about the radio observation and data reduction
are given in Gruppioni et al. (1997), where catalogs of 5$\sigma_{local}$
radio sources are given. The radio limits are about 0.2 mJy at both
frequencies. Cross--correlation of
the entire radio and X-ray catalogs has produced four positional coincidences
with a maximum difference of $\sim$10\arcs between radio
and X-ray positions. All these distances are between 1.1 and 2.4 times the 
combined X-ray and radio positional error ($\epsilon$). Fifteen more 
radio -- X-ray pairs have distances smaller
than 90\arcs, but none has a distance smaller than 25\arcs. For all 
these pairs the distances between radio and X-ray positions are larger than 
five times the combined positional error and can
therefore be considered random coincidences.
>From the number of these random pairs we estimate that the expected number of
similarly random pairs within 10\arcs is 0.20 $\pm$ 0.05.
If, instead of using a radius of 10\arcs,
which can be considered an ``a posteriori'' choice, we use for each X-ray
source a radius corresponding to the 95\% error circle, the estimated
number of random pairs inside the 95\% area becomes 0.40. On this basis we
conclude that probably none and at most one of the observed radio -- X-ray
coincidences is not real.
The four radio -- X-ray pairs correspond to a percentage of radio
detection of 8$\pm$4\% for our X-ray sources at a radio limit
of $\sim$ 0.2 mJy. 
Of the same order (10$\pm$5\%) is the percentage of X-ray detections for 
our sample of radio sources. Both these percentages
are in good agreement with what has been found by De Ruiter et al.
(1996) from VLA observations in the region of the Lockman Hole.
Table 2 lists these four radio -- X-ray pairs, where the columns are
X-ray number and flux from Table 1, radio number and flux from the 20 cm
catalog in Gruppioni et al. 1997, difference in position ($\Delta$) in arcsec 
and normalized to the combined X-ray and radio error ($\Delta$/$\varepsilon$). 
Three of these four radio -- X-ray pairs have ben optically identified 
(see Section 3.2). For all of them the offset between the radio and the
optical position of the suggested identification is less than 1.8\arcs, 
consistent with the combined radio and optical positional errors.
\begin{table}[t]
\caption[]{ The Radio -- X-ray Coincidences}
\begin{tabular}{rc|rc|cc} 
\hline\noalign{\smallskip}
\multicolumn{2}{c|}{X-ray~~~~} & \multicolumn{2}{c|}{Radio~~} & \multicolumn{2}{c}{Distance} \\
    \#~~~~~~& flux         &      \#     & flux        &$\Delta$&$\Delta$/$\varepsilon$\\
\multicolumn{2}{r|}{\hspace{1.2cm}[$\times 10^{-14}$ cgs]} &\hspace{0.8cm}&~~~[mJy]~~~& [arcsec] &\hspace{0.8cm}\\
\noalign{\smallskip}
\hline\noalign{\smallskip}
 ~~X013--01 &  5.00  &   15      & 158       &  3.8    &  2.0   \\
 ~~X021--05 &  2.36  &   38      &  1.25     &  2.3    &  1.1   \\
 ~~X409--28 &  0.61  &  ~~30     &  6.32     &  8.2    &  1.5   \\
 ~~X408--30 &  0.59  &  ~~35     &  0.41     & 10.3    &  2.4   \\
\noalign{\smallskip}\hline
\end{tabular}
\end{table}

\subsection{ The Optical Data}
In addition to the ESO 3.6m plates (U, J, F bands) which were already
available and have been used in the past to obtain a complete sample
of optically selected AGNs with $m_B \le 22.0$ (Zitelli et al. 1992),
in the years 1992--1994 we have obtained a set of U, B, V, R
CCD images at the ESO NTT. The V and R images cover $\sim$ 90\% of the circle
with 15\arcm \ radius and contain all but one of the X-ray sources within this
radius; the U and B images, which have a smaller field of view,
cover a smaller fraction of the 15\arcm~field, but still contain 45 out of
50 X-ray sources. Details about the observations, data reduction and
optical catalogs obtained from these CCD observations will be given
elsewhere (Mignoli et al. in preparation; see also Mignoli 1997).

The CCD limiting magnitudes vary
from field to field, but typically are of the order of 23.5 in U, 25.0
in B, 24.0 in V and R. The typical surface density of objects
at these limits is $\sim$ 60,000 per square degree.
Since the total area covered by the 2$\sigma$~(3$\sigma$) error circles of the
50 X-ray sources corresponds to 5.4~(10.4) sq.arcmin., this implies a total 
expected number of $\sim$~90~(175) catalogued objects inside the X-ray error 
boxes. 
In Table 3 we report the optical data for all the objects in our optical 
catalogs within a 3$\sigma$ error box, plus a few interesting objects
at slightly larger distance. 
(We recall that, under the assumption that the distribution function of the
positional errors follows a circular normal distribution, the 1$\sigma$,
2$\sigma$ and 3$\sigma$ error boxes correspond to radii equal to 1.51, 2.45
and 3.4 times the error in each coordinate, respectively.)
The first ten columns in the
Table give the X-ray number, the distance in right ascension and declination
between the X-ray source and the optical objects (arcsec), the total distance
both in arcsec and normalized to the X-ray positional error (\errx)
as defined in Section 2.1, \hbox{optical B} and R magnitudes and \hbox{U-B},
\hbox{B-V}, \hbox{V-R} colours and a morphological classification
(p for point-like source, e for extended, p/e or e/p for sources classified
differently in the blue and red bands). The morphological classification
is reliable only for objects which are more than $\sim$ 1.5 magnitudes
brighter than the limiting magnitude (Flynn, Gould and Bahcall 1996).
An asterisk in the magnitude columns
means that the magnitudes have been measured from the plates and then
converted to the Johnson system (see Gruppioni, Mignoli and Zamorani 1999). 

The next two columns give the likelihood
ratio (LR) computed from the B and R data. These LR values have been computed
following the procedure described by Sutherland and Saunders (1992), which,
differently from previous formulations, has been shown to be valid also
in the case of multiple candidates in the error boxes. For each optical
object its likelihood ratio is defined as:

\begin{equation}
LR = \frac{q(\rm{m,c})\; \em{f}(\rm{x,y})}{n(\rm{m,c})}
\end{equation}

\noindent
where $f$(x,y) is the probability distribution function of the positional 
errors, $n$(m,c) is the surface density of ``background'' objects with 
magnitude m and type c and $q$(m,c) is the probability distribution function
in magnitude and type of the optical counterparts. With this definition LR
is the ratio between the probability of finding the true optical counterpart
with the observed offset (x,y) from the X-ray position and the observed
magnitude and the probability of finding a similar chance background
object (see Eq. 1 and related discussion in Sutherland and Saunders 1992).
Assuming that the distribution function of the positional errors is
gaussian,

\begin{equation}
f(x,y) = \frac {  e^{- \frac {(x^2 + y^2)}{2 \sigma^2} }  } {2 \pi \sigma^2}
\end{equation}

\noindent
where x and y are the offsets in right ascension and declination between the
optical and X-ray sources and $\sigma$ is the positional error in each 
coordinate (see column 4 in Table 1).
We considered two different types of optical objects, i.e.
point--like and extended. For each of them the observed
$n$(m) has been obtained from our 
own CCD data, while $q$(m) has been estimated from the magnitude distribution
of the excess of objects in the 50 X-ray error boxes with respect to the
expected number of ``background'' objects.

The last two columns
give the redshift and the classification of the objects based on our
spectroscopic data. Objects which we consider to be the correct identification
are written in bold face in the last column. When the spectroscopic data
suggest identification with a group or cluster of galaxies, all the galaxies
at the redshift of the cluster are shown in bold face.

Figure 4 shows the finding charts for all the X-ray sources within 15\arcm,
ordered by decreasing X-ray flux, and, when available, the spectrum of the
most likely identification. For completeness we show here also the spectra
of the previously known, optically selected AGNs already published in
Zitelli et al. (1992). For all sources the optical images 
(1\arcm$\times$1\arcm) are taken from R CCDs, except for sources X046--03,
X240--09 and X045--27, for which  the images from the F plate are shown.
The two circles drawn on the figures correspond to the 68\% and 95\% error
boxes as defined above. The total number of objects in the 95\%
error boxes is $\sim$~140. Spectroscopy
for such a large number of faint objects was obviously not possible, and
therefore we had to decide a strategy for the follow--up spectroscopic 
observations of the most promising candidates. Our adopted strategy was the 
following:

i. First we have cross--correlated the positions of the 50 X-ray sources
with those of the 29 optically selected AGNs previously known in the same
area, finding 19 positional coincidences (see Section 2.1). 
With one exception (X019--17), all these AGNs have at least one of the two LR
values higher than 1.6. We have considered them as
likely identifications (these objects are indicated as {\bf AGN$^{(1)}$} 
in the last column of Table 3) and we decided not to
observe spectroscopically other objects in these error boxes, including the
X-ray source X019--17 in which the AGN has a relatively low likelihood ratio. We
note here that our LR values are based only on optical magnitudes and
morphology and do not take into account the spectroscopic information. 
At the typical magnitudes of the optical counterparts identified with AGNs
(21 $\le m_B \le$ 23) the ratio between the number of objects classified
as point--like objects in our CCD data and that of broad--line AGNs
(Mignoli and Zamorani 1998) is in the range 10--15. Therefore, the 
``a posteriori'' LR, when an AGN is found spectroscopically in any given
error box, would be about 
10--15 times higher than that listed in Table 3. 

ii. Since AGNs are well known to be the dominant optical counterpart of
faint X-ray sources, at least for $S_x \ge 5 \times 10^{-15} $ \erg,
we have then searched our optical catalogs for all the stellar
or slightly fuzzy objects within the remaining 31 X-ray error boxes.
We have then taken spectra for these objects helping us in making a priority
list in each error box with additional information from the colours, the 
magnitude and the distance between the objects and the X--centroid (i.e.
the LR value). In addition to
the objects that on the basis of these data were considered to be the most
likely AGN candidates, we gave high priority also to objects which 
coincide with a radio source (see Table 2) and to objects with colours
typical of M stars, which are the dominant spectral type of stars found in
faint X-ray surveys. M stars with the highest ratio between X-ray and
optical fluxes and with an X-ray flux equal to our limiting flux are
expected to have at most an optical magnitude of the order of $m_V \sim 19.75
\pm 0.25$ (see Maccacaro et al. 1988),
corresponding to $m_B \sim 21.25 \pm 0.25$. From our own
colour -- colour diagrams (see also Marano et al. 1988) we estimate that the
surface density of M stars with this limiting magnitude is $\sim$ 380 per square
degree, in good agreement (within 15\%) with the predictions
of the Bahcall and Soneira model for the structure of the Galaxy (see,
for example, Ratnatunga and Bahcall 1985).
Following this recipe, we found spectroscopically four stars (two M stars
with $m_b \sim 19.5$ (X030--15 and X028--33), and two very bright
F stars with $m_b \sim 10$ (X046--03 and X045--27)), 
eight broad-line AGNs (X012--02, X042--11, X040--32, X032--36, X251--37, 
X024--38, X015--39,  X306--42) and two radio galaxies (X021--05 and X409--28), 
one of which with broad MgII line. The blue magnitudes
of the AGNs and the two radio galaxies are in the range $21.7 - 23.7$. 

We also took a spectrum for the relatively bright object ($m_B$ = 21.44), with
LR values greater than 5.5, which is right at the center of the error box of 
source X051--43. Its spectrum does
not show any clear evidence for convincing emission or absorption features,
so that, even if we do consider it as the likely counterpart
of the X-ray source, we have not been able to spectroscopically classify it.
The absence of features in the spectrum suggests that it might be a BL Lac 
object. If so, it would be a somewhat anomalous BL Lac object. In fact, being 
not detected in the radio at a flux limit of about 0.2 mJy, it would be 
classified as a radio quiet BL Lac, with a radio to optical spectral index 
$\alpha_{ro} <$0.26. None of the X-ray selected BL Lac objects in the Einstein
Medium Sensitivity Survey (Stocke et al. 1991), and just a few in the much 
larger samples of ROSAT selected BL Lacs (see, for example, Perlman et al. 1998)
have such a low value of $\alpha_{ro}$. 

\begin{table*}             
\caption[]{ The Optical Data}
\begin{tabular}{rrrrrrrrrcrrll}
\hline\hline
\\
 ID.  & $\Delta\alpha$ & $\Delta\delta$ & $\Delta$\arcs/$\displaystyle{\Delta\over\varepsilon_{\rm{x}}}$ &
 \multicolumn{1}{c}{~m$_B$} & \multicolumn{1}{c}{m$_R$} & U-B & B-V & V-R & &
 \multicolumn{1}{c}{$LR_B$} &\multicolumn{1}{c}{$LR_R$} & \multicolumn{1}{c}{z} & object notes\\
 &\multicolumn{1}{c}{~~[\arcs]}  & \multicolumn{1}{c}{~~[\arcs]} \\
\hline     
 X013--01 &   1.2 &  -0.0 &  1.2/0.8 & $>$25.00  &    23.17  &         &          & $>$ 1.13 & e &       &  4.73 & & \\
          &  -4.0 &  -1.7 &  4.3/2.8 &    19.96  &    19.61  &   -0.55 &    -0.09 &     0.44 & p &  1.25 &  1.63 & 1.663 & {\bf AGN$^{(1)}$}\\
\hline    
 X012--02 &  -7.6 &   0.2 &  7.6/2.8 &    21.94  &    20.58  &   -0.70 &     0.63 &     0.73 & p &  1.16 &  0.65 & 1.378 & {\bf AGN}\\
\hline    
 X046--03 &  -1.4 &   0.9 &  1.6/1.0 & $\sim$9.3 &           &         & $\sim$0.5&          &   &       &       & 0.00  & {\bf F6IV~star}\\
\hline    
 X036--04 &  -0.4 &  -0.3 &  0.5/0.2 &    18.04* &    17.07  &   -0.78 &     0.55 &     0.42 & p & 17.86 & 28.60 & 2.531 & {\bf AGN$^{(1)}$}\\
          &   0.8 &   5.1 &  5.1/2.1 & $>$23.50* &    23.44  &         & $>$-0.67 &     0.73 & e &       &  0.29 & & \\
\hline    
 X021--05 &  -2.7 &   0.0 &  2.7/1.5 &    21.79  &    19.28  &    0.66 &     1.48 &     1.03 & e &  6.83 &  3.58 & 0.387 & {\bf RadioGal}\\
\hline    
 X025--06 &   1.0 &  -1.6 &  1.9/1.0 &    21.43  &    20.97  &   -0.52 &     0.36 &     0.10 & p & 71.44 & 45.19 & 0.808 & {\bf AGN$^{(1)}$}\\
\hline    
 X027--07 &   3.0 &   3.3 &  4.4/2.0 &    21.17  &    20.66  &   -0.55 &     0.19 &     0.32 & p & 11.89 &  7.52 & 0.636 & {\bf AGN$^{(1)}$}\\
          &  -3.4 &  -5.6 &  6.6/2.9 &    23.34  &    22.87  &   -0.86 &     0.29 &     0.18 & e &  0.12 &  0.09 & & \\
\hline    
 X041--08 &   1.8 &   3.6 &  4.0/1.5 &    21.62  &    21.06  &   -0.67 &     0.35 &     0.21 & p & 20.74 & 10.78 & 2.161 & {\bf AGN$^{(1)}$}\\
          &  -3.5 &  -5.7 &  6.7/2.6 &    24.04  &    23.46  &   -0.66 &     0.41 &     0.17 & e &  0.08 &  0.08 & & \\
          &   0.3 &  -7.5 &  7.5/2.9 &    24.20  & $>$23.50  &$>$-0.70 &     0.48 & $<$ 0.22 & e &  0.04 &       & & \\
          &  -8.5 &   2.3 &  8.8/3.4 &    24.15  & $>$23.50  &$>$-0.65 &    -0.21 & $<$ 0.86 & e &  0.01 &       & & \\
\hline    
 X240--09 &  -0.3 &   4.0 &  4.0/1.0 &    21.81* &    20.87* &   -0.71 &     0.52 &     0.42 & p & 17.75 &  9.99 & 0.854 & {\bf AGN$^{(1)}$}\\
          &  12.4 &  -1.9 & 12.5/3.2 &    22.54* &    21.91* &   -0.74 &     0.34 &     0.29 & e &  0.01 &  0.02 & & \\
\hline    
 X033--10 &   7.2 &   1.4 &  7.3/1.7 &    21.90* &    20.91  &   -0.90 &     0.65 &     0.34 & p &  5.70 &  3.21 & 0.983 & {\bf AGN$^{(1)}$}\\
 &   1.0 &  12.6 & 12.6/3.0 & $>$23.50* &    21.86  &         & $>$ 0.72 &     0.92 & p &       &  0.09 & & \\
\hline    
 X042--11 &  -0.8 &   2.4 &  2.5/1.1 &    22.99  &    22.04  &   -0.33 &     0.58 &     0.37 &p/e& 19.50 &  5.43 & 1.062 & {\bf AGN}\\
 &  -6.8 &  -3.8 &  7.8/3.3 &   $-out-$ &    22.10  &         &          &     1.00 & e &       &  0.04 & & \\
\hline    
 X043--12 &   1.6 &   3.1 &  3.5/1.5 &    23.80  &    22.61  &   -0.32 &     0.17 &     1.02 &e/p&  1.66 &  0.87 & 2.80: & {\bf AGN2}~(?)\\
 &   5.3 &  -5.8 &  7.9/3.4 &    24.15  &    22.93  &   -0.47 &     0.95 &     0.27 & e &  0.01 &  0.02 & & \\
\hline    
 X023--13 &   0.3 &  -0.7 &  0.8/0.3 &    22.24  &    21.99  &   -0.45 &    -0.10 &     0.35 & p & 36.47 & 19.86 & 1.573 & {\bf AGN$^{(1)}$}\\
 &   4.5 &  -8.1 &  9.3/3.4 &    23.94  &    21.32  &$>$-0.44 &     1.42 &     1.20 & e &  0.01 &  0.02 & & \\
\hline    
 X108--14 &   3.6 &   0.6 &  3.6/0.8 &    22.42  &    21.63  &   -1.55 &     0.23 &     0.56 & p & 10.04 &  5.47 & 1.374 & {\bf AGN$^{(1)}$}\\
 &  -4.8 &   3.1 &  5.7/1.3 &    21.30  &    20.36  &    0.05 &     0.52 &     0.42 & e &  2.90 &  1.71 & & \\
\hline    
 X030--15 &   0.5 &   3.5 &  3.6/0.8 &    24.07  & $>$23.50  &   -0.35 & $<$-0.23 &          & p &  0.60 &       & & \\
 &   0.5 &   7.6 &  7.6/1.8 &    19.28  &    16.86  &    1.20 &     1.59 &     0.83 & p &  1.95 &  2.59 & 0.00  & {\bf M~star}\\
 &  -8.1 &  10.1 & 13.0/3.0 &    23.29  &    22.20  &   -0.11 &     0.87 &     0.22 & e &  0.02 &  0.03 & & \\
 & -13.0 &  -5.7 & 14.2/3.3 &    22.99  &    22.24  &   -0.86 &     0.33 &     0.42 & e &  0.01 &  0.01 & & \\
\hline    
 X304--16 &  -1.0 &   1.6 &  1.9/0.6 &    21.53  &    20.73  &   -0.74 &     0.39 &     0.41 & p & 39.05 & 21.97 & 1.192 & {\bf AGN$^{(1)}$}\\
 &  -2.6 &   5.7 &  6.3/2.0 &    24.04  &    23.05  &   -0.84 &     0.40 &     0.59 & e &  0.21 &  0.21 & & \\
 &   9.8 &  -2.2 & 10.1/3.2 &    23.59  &    22.27  &   -0.70 &     0.46 &     0.86 & e &  0.02 &  0.03 & & \\
\hline    
 X019--17 &   0.9 &  -2.8 &  2.9/0.8 &    24.34  &    23.47  &   -0.90 &     0.78 &     0.09 & e &  0.85 &  0.85 & & \\
 &  -2.8 &   8.7 &  9.2/2.5 &    23.45  &    21.20  &$>$ 0.05 &     1.35 &     0.90 & p &  0.30 &  0.74 & & \\
 &  -5.5 & -10.4 & 11.7/3.2 &    22.40* &    21.69  &   -0.70 &     0.49 &     0.22 &p/e&  0.12 &  0.03 & 0.614 & {\bf AGN$^{(1)}$}\\
\hline    
 X029--18 &   3.5 &   4.8 &  6.0/1.6 &    23.67  &    21.94  &   -0.10 &     1.11 &     0.62 & e &  0.56 &  1.20 & & \\
 &   0.9 &  -6.2 &  6.2/1.6 &    21.66  &    21.15  &   -1.01 &     0.23 &     0.28 & p &  8.42 &  4.37 & 1.254 & {\bf AGN$^{(1)}$}\\
\hline    
 X039--19 & -10.5 &   2.2 & 10.7/3.3 &    23.63  &    22.41  &   -0.20 &     0.99 &     0.23 & e &  0.01 &  0.02 & ?.??? & low S/N\hfill(A)\\
\hline    
 X001--20 &  -5.4 &  -1.8 &  5.7/1.6 &    20.39  &    19.71  &   -0.83 &     0.26 &     0.42 & p &  2.31 &  3.92 & 1.353 & {\bf AGN$^{(1)}$}\\
 &  -8.4 &   7.0 & 10.9/3.0 &    22.72  &    22.73  &   -1.00 &    -0.09 &     0.08 & p &  0.16 &  0.01 & & \\
\hline    
 X049--21 &  -3.2 &  -1.7 &  3.6/1.2 &    23.86  &    22.36  &   -0.61 &     0.68 &     0.82 & e &  1.50 &  2.82 & ?.??? & low S/N\\
\hline    
 X211--22 &  -0.2 &  -6.9 &  6.9/1.6 &    24.27  &    23.31  &   -0.51 & $<$-0.03 & $>$ 0.99 & e &  0.23 &  0.23 & & \\
 &   9.3 &   7.5 & 12.0/2.8 &    20.97  &    19.35  &    0.11 &     1.06 &     0.56 & e &  0.07 &  0.04 & 0.180 & early gal\\
 & -12.5 &   0.5 & 12.5/2.9 &    22.20  &    20.26  &   -0.45 &     0.86 &     1.08 &p/e&  0.23 &  0.07 & 0.281 & {\bf AGN (Sy 1)}\\
 & -12.3 &  -4.0 & 13.0/3.0 &    24.37  &    22.99  &   -0.16 &     0.43 &     0.95 & e &  0.01 &  0.02 & & \\
 & -13.5 &   0.1 & 13.5/3.1 &    22.12  &    21.36  &   -0.04 &     0.41 &     0.35 & p &  0.12 &  0.10 & 0.00  & B~star\\
 &   7.0 &  12.9 & 14.7/3.4 &    24.60  &    23.49  &$>$-1.10 &     0.64 &     0.47 & e &  0.00 &  0.00 & & \\
\hline
\end{tabular}
\end{table*}
\begin{table*}              
\begin{tabular}{rrrrrrrrrcrrll}
\hline
\\
 ID.  & $\Delta\alpha$ & $\Delta\delta$ & $\Delta$\arcs/$\displaystyle{\Delta\over\varepsilon_{\rm{x}}}$ &
 \multicolumn{1}{c}{~m$_B$} & \multicolumn{1}{c}{m$_R$} & U-B & B-V & V-R & &
 \multicolumn{1}{c}{$LR_B$} &\multicolumn{1}{c}{$LR_R$} & \multicolumn{1}{c}{z} & object notes\\
 &\multicolumn{1}{c}{~~[\arcs]}  & \multicolumn{1}{c}{~~[\arcs]} \\
\hline    
 X404--23 &  11.8 &  -1.4 & 11.9/1.3 &    24.03  &    22.74  &$>$-0.53 &     0.70 &     0.59 & e &  0.08 &  0.17 & & \\
 &   6.0 &  10.9 & 12.4/1.3 & $>$25.00  &    23.25  &         &          & $>$ 1.05 & e &       &  0.07 & & \\
 &  -9.8 & -10.1 & 14.1/1.5 &    23.67  &    22.54  &   -0.38 &     0.87 &     0.26 & e &  0.10 &  0.12 & & \\
 & -16.1 &  -1.6 & 16.2/1.7 &    24.25  &    22.67  &$>$-0.75 &     0.76 &     0.82 & e &  0.04 &  0.09 & & \\
 &   6.2 & -16.5 & 17.6/1.9 &    24.47  &    23.41  &$>$-0.97 &     0.26 &     0.80 & e &  0.03 &  0.03 & & \\
 &   8.9 & -16.5 & 18.8/2.0 &    24.89  &    23.18  &$>$-1.39 &     0.70 &     1.01 & e &  0.01 &  0.02 & & \\
 & -12.9 & -15.5 & 20.2/2.2 &    23.40  &    22.34  &   -0.77 &     0.57 &     0.49 & e &  0.05 &  0.06 & 0.204 & NELG\hfill(G)\\
 & -18.0 &  -9.3 & 20.3/2.2 &    24.23  &    23.60  &   -0.17 &     0.16 &     0.47 & e &  0.02 &  0.01 & & \\
 &  -3.7 &  21.3 & 21.6/2.3 &    24.32  &    23.47  &   -0.66 &     0.41 &     0.44 & e &  0.01 &  0.01 & & \\
 &  11.9 & -19.3 & 22.6/2.4 &    23.55  &    21.01  &$>$-0.05 &     1.31 &     1.23 & p &  0.03 &  0.14 & 0.00  & M~star\hfill(J)\\
 &  23.3 &  14.2 & 27.3/2.9 &    23.65  &    21.81  &$>$-0.15 &     0.85 &     0.99 & e &  0.00 &  0.01 & & \\
 &  24.6 & -12.6 & 27.7/3.0 &    23.80  &    22.20  &   -1.18 &     0.75 &     0.85 & e &  0.00 &  0.01 & & \\
 &  24.1 &  13.7 & 27.7/3.0 &    23.82  & $>$23.50  &$>$-0.32 &    -0.24 & $<$ 0.56 & e &  0.00 &       & & \\
 &  13.6 & -24.3 & 27.8/3.0 &    23.67  &    23.04  &   -0.67 &    -0.16 &     0.79 & e &  0.00 &  0.00 & & \\
 &  27.8 &  -5.3 & 28.3/3.0 &    22.51  &    21.70  &   -0.36 &     0.52 &     0.29 & e &  0.00 &  0.01 & 0.157 & Sy2~(?) \hfill(N)\\
 & -27.1 &   8.9 & 28.5/3.1 &    23.76  &    21.43  &$>$-0.26 &     1.28 &     1.05 & p &  0.01 &  0.02 & & \\
 & -15.7 &  24.5 & 29.1/3.1 & $>$25.00  &    23.35  &         & $>$ 0.77 &     0.88 & e &       &  0.00 & & \\
 &   8.3 &  28.2 & 29.4/3.2 &    24.45  &    21.94  &$>$-0.95 &     1.51 &     1.00 & p &  0.00 &  0.01 & & \\
\hline    
 X031--24 &   0.0 &   2.8 &  2.8/0.6 &    21.37* &    20.71  &   -0.72 &     0.32 &     0.34 & p & 17.50 & 11.07 & 0.409 & {\bf AGN$^{(1)}$}\\
 &  -1.3 &  -3.7 &  3.9/0.9 & $>$23.50* &    21.23  &         & $>$ 1.19 &     1.08 & e &       &  1.90 & & \\
 &   5.8 &   1.0 &  5.9/1.4 & $>$23.50* &    23.18  &         & $>$-0.77 &     1.09 & e &       &  0.33 & & \\
 &  -0.9 &  -7.8 &  7.8/1.8 & $>$23.50* &    21.71  &         & $>$ 1.37 &     0.42 & e &       &  0.64 & & \\
 &   0.1 & -13.6 & 13.6/3.1 & $>$23.50* &    22.82  &         & $>$ 0.21 &     0.47 & e &       &  0.01 & & \\
\hline    
 X050--25 &  -0.4 &   0.7 &  0.8/0.2 &    20.77  &    20.39  &   -0.72 &     0.04 &     0.34 & p & 18.47 & 13.93 & 1.315 & {\bf AGN$^{(1)}$}\\
 &   4.0 &  -5.8 &  7.1/1.9 &    24.17  &    23.08  &$>$-0.67 &     0.31 &     0.78 & e &  0.19 &  0.19 & & \\
\hline    
 X235--26 &  -0.9 &  -1.7 &  1.9/0.4 &    21.40  &    20.96  &   -0.00 &     0.26 &     0.18 & p & 13.94 &  8.82 & 2.536 & {\bf AGN$^{(1)}$}\\
 &   2.5 &  -1.0 &  2.7/0.5 &    24.92  &    23.42  &$>$-1.42 & $<$ 0.62 & $>$ 0.88 & e &  0.23 &  0.50 & & \\
 &  -0.5 &  -5.2 &  5.2/1.0 &    22.28  &    20.88  &   -0.42 &     0.69 &     0.71 & e &  1.54 &  1.24 & & \\
 &   5.5 &  -9.5 & 11.0/2.1 &    23.37  &    22.14  &   -0.62 &     0.57 &     0.66 & e &  0.19 &  0.21 & & \\
 &   2.7 &  10.9 & 11.2/2.1 &    24.33  &    23.37  &$>$-0.83 &     0.40 &     0.56 & e &  0.06 &  0.06 & & \\
 & -10.5 &  -9.7 & 14.3/2.7 &    24.46  &    23.73  &   -0.64 &     0.03 &     0.70 & e &  0.01 &  0.01 & & \\
 &   7.6 &  14.9 & 16.8/3.2 & $>$25.00  &    22.55  &         & $>$ 0.88 &     1.57 & e &       &  0.01 & & \\
 &  11.6 & -13.4 & 17.7/3.4 &    23.45  &    22.10  &    0.08 &     0.49 &     0.86 & e &  0.01 &  0.01 & & \\
\hline    
 X045--27 &   2.6 &  -9.8 & 10.1/2.9 &$\sim$10.0 &           &         & $\sim$0.3&          &   &       &       & 0.00  & {\bf F5V~star}\\
\hline    
 X409--28 &   0.1 &  -7.5 &  7.5/1.4 &    23.70  &    21.81  &   -0.65 &     0.97 &     0.92 & e &  0.36 &  0.79 & 0.957 & {\bf AGN (BLRG)}\\
 &  -6.5 &   5.0 &  8.2/1.6 &    24.67  & $>$23.50  &$>$-1.17 & $<$ 0.37 &          & e &  0.10 &       & & \\
 &  -9.0 &  -3.1 &  9.5/1.8 &    24.46  &    23.21  &   -0.78 &     0.24 &     1.01 & e &  0.11 &  0.11 & & \\
 &   3.5 &  -9.3 &  9.9/1.9 &    24.27  &    22.47  &$>$-0.77 &     0.87 &     0.93 & e &  0.09 &  0.32 & & \\
 &  -6.5 &  -8.6 & 10.7/2.0 & $>$25.00  &    22.76  &         &          & $>$ 1.54 & e &       &  0.15 & & \\
 & -10.1 &  -6.6 & 12.1/2.3 &    23.65  &    22.68  &   -0.66 &     0.41 &     0.56 & e &  0.07 &  0.08 & & \\
 &  13.5 &   5.6 & 14.6/2.8 &    24.41  &    23.50  &   -0.56 &     0.29 &     0.62 & e &  0.01 &  0.00 & & \\
\hline    
 X301--29 &  -2.5 &   0.4 &  2.6/0.6 &    21.13* &    20.41* &   -0.88 &     0.40 &     0.32 & p & 19.92 & 10.18 & 1.709 & {\bf AGN$^{(1)}$}\\
 &  -0.9 &   5.9 &  6.0/1.5 &    23.86  &    22.67  &   -0.42 &     0.56 &     0.63 & e &  0.57 &  0.68 & & \\
 &  -2.7 & -10.4 & 10.8/2.6 &    19.13* &    16.30* &    1.09 &     1.59 &     1.24 & p &  0.32 &  0.25 & & \\
 &   4.7 &  11.3 & 12.2/3.0 &    22.79  &    21.13  &   -1.38 &     0.96 &     0.70 & e &  0.03 &  0.04 & & \\
\hline    
 X408--30 &  -9.6 &   3.5 & 10.2/2.5 &    24.76  & $>$23.50  &$>$-1.26 &     0.40 & $<$ 0.86 & e &  0.02 &       & & \\
 &  12.8 &   5.9 & 14.1/3.4 &    24.56  &    23.00  &$>$-1.06 &     0.70 &     0.86 & e &  0.00 &  0.00 & & \\
 &   4.6 &  13.8 & 14.6/3.6 &    23.49  &    21.53  &   -0.30 &     1.04 &     0.92 & e &  0.00 &  0.01 & 0.814 & NELG\hfill(C)\\
\hline    
 X207--31 &  -5.7 &  -0.7 &  5.7/1.2 &    23.37  &    22.19  &   -0.38 &     0.64 &     0.54 & e &  0.98 &  1.11 & & \\
 &  -1.6 &  -8.0 &  8.1/1.7 &    23.78  &    22.43  &   -0.44 &     0.84 &     0.51 & e &  0.29 &  0.56 & 0.58: & {\bf cluster gal}\\
 & -10.0 &   4.8 & 11.1/2.3 &    24.05  &    22.97  &$>$-0.55 &     0.44 &     0.64 & e &  0.05 &  0.11 & & \\
 &  -6.7 &  10.0 & 12.0/2.5 &    24.58  &    23.55  &$>$-1.08 &     0.45 &     0.58 & e &  0.02 &  0.02 & & \\
 & -12.3 &   1.0 & 12.3/2.5 & $>$25.00  &    23.22  &         &          & $>$ 1.08 & e &       &  0.03 & & \\
 & -14.0 &  -5.8 & 15.2/3.1 &    24.58  & $>$23.50  &$>$-1.08 & $<$ 0.28 &          & e &  0.00 &       & & \\
 & -15.7 &   6.4 & 16.9/3.5 &    23.94  &    22.38  &   -0.43 &     0.57 &     0.99 & e &  0.00 &  0.01 & & \\
 &  -5.8 &  17.3 & 18.2/3.7 &    24.34  &    23.69  &   -0.84 &     0.11 &     0.54 & e &  0.00 &  0.00 & & \\
 & -17.3 &  -6.3 & 18.5/3.8 &    22.68  &    19.65  &$>$ 0.82 &     1.72 &     1.31 & e &  0.00 &  0.00 & 0.584 & {\bf cD gal}\\
\hline
\end{tabular}
\end{table*}
\begin{table*}            
\begin{tabular}{rrrrrrrrrcrrll}
\hline
\\
 ID.  & $\Delta\alpha$ & $\Delta\delta$ & $\Delta$\arcs/$\displaystyle{\Delta\over\varepsilon_{\rm{x}}}$ &
 \multicolumn{1}{c}{~m$_B$} & \multicolumn{1}{c}{m$_R$} & U-B & B-V & V-R & &
 \multicolumn{1}{c}{$LR_B$} &\multicolumn{1}{c}{$LR_R$} & \multicolumn{1}{c}{z} & object notes\\
 &\multicolumn{1}{c}{~~[\arcs]}  & \multicolumn{1}{c}{~~[\arcs]} \\
\hline     
 X040--32 &   6.0 &   2.6 &  6.6/1.9 &    23.37  &    22.31  &   -0.68 &     0.57 &     0.49 & p &  1.30 &  0.99 & 1.204 & {\bf AGN}\\
 &  -6.1 &   4.8 &  7.8/2.2 &    24.53  & $>$23.50  &   -0.82 &    -0.30 & $<$ 1.33 & e &  0.08 &       & & \\
 &   8.3 &   1.5 &  8.4/2.4 &    23.26  &    21.99  &   -0.74 &     0.52 &     0.75 & e &  0.21 &  0.27 & 0.804 & NELG\\
 &   9.0 &   4.1 &  9.8/2.8 &    23.61  &    22.27  &    0.41 &     0.52 &     0.82 & e &  0.04 &  0.08 & & \\
\hline    
 X028--33 &  -3.1 &  -0.0 &  3.1/0.6 & $>$25.00  & $>$23.50  &$<$-1.77 &          &          & p &       &       & & \\
 &   4.3 &  -0.3 &  4.3/0.9 &    23.20  &    22.20  &   -0.78 &     0.40 &     0.60 & e &  1.35 &  1.53 & & \\
 &  -5.5 &  -1.8 &  5.8/1.2 &    19.16  &    17.22  &    1.06 &     1.29 &     0.65 & p &  3.64 &  3.59 & 0.00  & {\bf M~star}\\
 &  -1.8 & -10.7 & 10.8/2.3 &    21.95  &    20.98  &   -0.61 &     0.36 &     0.61 & e &  0.23 &  0.19 & 0.657 & NELG\\
 &  11.1 &  -4.4 & 11.9/2.5 &    23.88  &    23.32  &   -0.79 &     0.28 &     0.28 & e &  0.06 &  0.03 & & \\
 &  12.2 &   8.0 & 14.6/3.0 &    23.89  &    23.01  &   -0.43 &     0.10 &     0.78 & p &  0.02 &  0.01 & & \\
 &  16.0 &   1.8 & 16.1/3.4 &    24.25  &    21.71  &$>$-0.75 &     1.26 &     1.28 & e &  0.00 &  0.01 & & \\
\hline    
 X250--34 &   7.1 &   0.6 &  7.1/1.6 &    24.30  &    23.59  &$>$-0.80 & $<$ 0.00 & $>$ 0.71 & e &  0.22 &  0.16 & & \\
 &  -8.9 &   5.6 & 10.5/2.4 &    23.29  &    22.94  &$>$ 0.21 &     0.08 &     0.27 & e &  0.13 &  0.10 & & \\
 &  -9.1 &  -6.0 & 10.9/2.5 &    24.32  &    23.68  &$>$-0.82 & $<$ 0.02 & $>$ 0.62 & e &  0.04 &  0.02 & & \\
 &   6.8 &  -8.9 & 11.2/2.6 & $>$25.00  &    23.18  &         &          & $>$ 1.12 & e &       &  0.03 & & \\
\hline    
 X011--35 &   0.6 &  -0.8 &  1.0/0.2 &    21.94  &    20.33  &   -0.20 &     1.14 &     0.47 & e &  1.80 &  2.03 & 0.189 & abs.gal\hfill(A)\\
 &   1.7 &   2.5 &  3.1/0.5 &    24.49  &    23.75  &$>$-0.99 & $<$ 0.19 & $>$ 0.55 & e &  0.37 &  0.23 & & \\
 &   0.8 &   6.3 &  6.4/1.0 &    23.61  &    22.40  &   -0.32 &     0.63 &     0.58 & e &  0.43 &  0.81 & & \\
 &  -7.8 &   5.5 &  9.5/1.6 &    21.95  &    20.63  &   -0.67 &     0.76 &     0.56 & e &  0.54 &  0.44 & 0.391 & {\bf NELG}\hfill(D)\\
 &  -7.6 &  -5.8 &  9.6/1.6 &    23.99  & $>$23.50  &$>$-0.49 &    -0.19 & $<$ 0.68 & e &  0.22 &       & & \\
 &  -1.0 & -10.2 & 10.2/1.7 &    24.03  &    21.34  &$>$-0.53 &     1.34 &     1.35 & e &  0.10 &  0.36 & 0.586 & early gal\hfill(F)\\
 &   7.3 &  -7.4 & 10.4/1.7 &    24.02  &    23.42  &   -0.97 &    -0.05 &     0.65 & p &  0.10 &  0.10 & & \\
 & -11.9 &   0.2 & 11.9/1.9 &    23.63  &    22.94  &$>$-0.13 &     0.39 &     0.30 & e &  0.11 &  0.14 & & \\
 &  -6.0 & -12.0 & 13.4/2.2 &    22.52  &    21.29  &   -0.39 &     0.64 &     0.59 & e &  0.09 &  0.13 & ?.??? & low S/N\hfill(I)\\
 & -10.0 & -10.9 & 14.8/2.4 &    23.45  &    23.63  &$>$ 0.05 &     0.05 &    -0.23 & e &  0.07 &  0.02 & & \\
 &   7.4 & -16.9 & 18.5/3.0 &    24.72  &    22.17  &$>$-1.22 &     1.41 &     1.14 & p &  0.00 &  0.02 & & \\
 & -15.0 &  13.4 & 20.1/3.3 &    22.00  &    20.15  &$>$ 1.00 &     1.14 &     0.71 & e &  0.01 &  0.01 & 0.390 & {\bf early+[OII]}\hfill(L)\\
 &  -4.8 & -21.2 & 21.7/3.5 &    24.02  &    22.66  &$>$-0.52 &     0.90 &     0.46 & e &  0.00 &  0.00 & & \\
 & -19.5 &  10.9 & 22.3/3.6 &   $-out-$ &    22.86  &         &          &     0.93 & e &       &  0.00 & & \\
 &   8.9 & -22.3 & 24.0/3.9 &    22.90  &    20.57  &$>$ 0.60 &     1.31 &     1.02 & e &  0.00 &  0.00 & 0.390 & {\bf early gal}\hfill(O)\\ 
\hline    
 X032--36 &   5.3 &  -3.1 &  6.1/1.1 &    24.43  &    23.19  &$>$-0.93 &     0.51 &     0.73 & e &  0.28 &  0.28 & & \\
 &  -3.8 &   9.4 & 10.1/1.9 &    24.24  &    22.81  &$>$-0.74 &     0.87 &     0.56 & e &  0.09 &  0.20 & & \\
 &   4.4 &   9.2 & 10.3/1.9 &    21.26  &    20.61  &   -0.15 &     0.34 &     0.31 & p &  2.27 &  1.44 & 0.00  & B~star\\
 &   0.5 &  12.4 & 12.4/2.3 &    23.36  &    21.98  &   -0.56 &     0.79 &     0.59 & p &  0.54 &  0.37 & 1.190 & {\bf AGN}\\
 &  10.6 &   9.0 & 13.9/2.6 & $>$25.00  &    22.88  &         & $>$ 1.09 &     1.03 & e &       &  0.04 & & \\
 &  -0.2 & -14.3 & 14.3/2.6 &    24.21  &    23.46  &$>$-0.71 &     0.66 &     0.09 & p &  0.02 &  0.02 & & \\
 &  -8.9 & -13.4 & 16.1/3.0 &    24.21  &    23.43  &$>$-0.71 &     0.22 &     0.56 & e &  0.01 &  0.01 & & \\
 &  16.3 &   1.6 & 16.4/3.0 & $>$25.00  &    23.27  &         &          & $>$ 1.03 & e &       &  0.01 & & \\
\hline    
 X251--37 &  -6.1 &   4.5 &  7.6/1.3 &    21.72  &    20.68  &   -0.20 &     0.70 &     0.34 & p &  5.70 &  3.20 & 2.710 & {\bf AGN}\\
 &   4.3 &   6.9 &  8.1/1.3 &    24.34  &    23.40  &$>$-0.84 &     0.57 &     0.37 & e &  0.17 &  0.17 & & \\
 &   3.5 &  12.4 & 12.9/2.1 &    24.61  & $>$23.50  &$>$-1.11 & $<$ 0.31 &          & e &  0.03 &       & & \\
 & -13.7 &   0.8 & 13.7/2.3 &    20.03  &    18.40  &    0.28 &     1.07 &     0.56 & e &  0.23 &  0.14 & 0.091 & early gal\\
 & -12.3 &   8.0 & 14.7/2.4 &    23.95  &    22.80  &$>$-0.45 &     0.81 &     0.34 & e &  0.04 &  0.05 & & \\
\hline    
 X024--38 &   0.8 &  -0.0 &  0.8/0.1 &    22.19  &    20.76  &   -0.37 &     0.77 &     0.66 & p &  9.29 &  8.72 & 1.430 & {\bf AGN}\\
 &   3.6 &  -3.6 &  5.1/0.9 &    17.94  &    16.82  &    0.02 &     0.45 &     0.67 & p &  3.39 &  5.06 & 0.00  & B~star\\
 &   0.4 &   9.3 &  9.3/1.7 &    23.21  &    22.23  &   -0.24 &     0.73 &     0.25 & p &  0.75 &  0.57 & 0.276 & NELG\\
 &   8.5 &  10.1 & 13.2/2.4 &    22.76  &    21.26  &    0.24 &     0.88 &     0.62 & e &  0.06 &  0.09 & & \\
 &   2.4 &  15.8 & 16.0/3.0 &    23.82  &    23.54  &$>$-0.32 &     0.50 &    -0.22 & e &  0.01 &  0.01 & & \\
 & -17.4 &  -1.8 & 17.5/3.2 &    24.35  &    22.65  &$>$-0.85 &     0.95 &     0.75 & e &  0.00 &  0.01 & & \\
 &  17.8 &  -2.1 & 17.9/3.3 &    14.59  &    13.36  &   -0.54 &     0.23 &     1.00 & p &  0.15 &  0.06 & 0.00  & A~star\\
 & -19.1 &  -1.1 & 19.1/3.5 &    23.95  & $>$23.50  &$>$-0.45 &    -0.34 & $<$ 0.79 & e &  0.00 &       & & \\
 & -10.1 &  16.9 & 19.7/3.6 &    16.99  &    15.68  &    0.29 &     0.42 &     0.89 & p &  0.00 &  0.00 & 0.00  & G~star\\
\hline
 X015--39 &  -5.4 &  -5.8 &  7.9/1.1 &    23.27  &    22.01  &   -0.46 &     0.69 &     0.57 & p &  1.04 &  0.79 & 0.500 & {\bf AGN (Sy 1)}\\
 &  13.1 &  -8.1 & 15.4/2.2 & $>$25.00  &    23.22  &         &          & $>$ 1.08 & e &       &  0.03 & & \\
 & -17.4 &  -2.8 & 17.6/2.5 &    23.48  &    22.73  &   -0.68 &     0.15 &     0.60 & e &  0.04 &  0.03 & & \\
 &   7.2 & -17.2 & 18.7/2.7 &    24.64  &    23.43  &$>$-1.14 & $<$ 0.34 & $>$ 0.87 & e &  0.01 &  0.01 & & \\
 & -17.3 &  -7.5 & 18.8/2.7 &    23.85  &    22.62  &   -0.79 &     0.21 &     1.02 & e &  0.01 &  0.02 & ?.??? & low S/N\\
\hline    
\end{tabular}
\end{table*}
\begin{table*}                 
\begin{tabular}{rrrrrrrrrcrrll}
\hline
\\
 ID.  & $\Delta\alpha$ & $\Delta\delta$ & $\Delta$\arcs/$\displaystyle{\Delta\over\varepsilon_{\rm{x}}}$ &
 \multicolumn{1}{c}{~m$_B$} & \multicolumn{1}{c}{m$_R$} & U-B & B-V & V-R & &
 \multicolumn{1}{c}{$LR_B$} &\multicolumn{1}{c}{$LR_R$} & \multicolumn{1}{c}{z} & object notes\\
 &\multicolumn{1}{c}{~~[\arcs]}  & \multicolumn{1}{c}{~~[\arcs]} \\
\hline     
 X236--40 &   5.6 &  -0.3 &  5.6/1.3 &    22.23* &    21.64  &   -1.00 &     0.24 &     0.35 & p &  6.38 &  3.47 & 1.140 & {\bf AGN$^{(1)}$}\\
 &  -7.7 &   0.5 &  7.7/1.8 & $>$23.50* &    22.00  &         & $>$ 0.11 &     1.39 & e &       &  0.57 & & \\
 &  -8.7 &  -2.1 &  8.9/2.1 &    24.55  &    23.73  &$>$-1.05 &     0.19 &     0.63 & e &  0.07 &  0.06 & & \\
 & -12.0 &  -3.0 & 12.4/2.9 & $>$25.00  &    23.19  &         & $>$ 0.39 &     1.42 & e &       &  0.01 & & \\
 &  -6.2 &  12.3 & 13.8/3.2 &    24.30  &    22.64  &   -0.60 &     0.65 &     1.01 & e &  0.00 &  0.01 & & \\
\hline    
 X234--41 &   0.9 &   1.2 &  1.5/0.3 &    23.38  &    22.18  &    0.17 &     0.77 &     0.43 & e &  1.58 &  1.78 & & \\
 &  -0.5 &   4.1 &  4.1/0.8 &    23.83  &    23.29  &   -0.54 &    -0.03 &     0.57 & e &  0.73 &  0.41 & & \\
 &   1.5 &  -4.1 &  4.3/0.8 &    23.72  &    22.58  &    0.41 &     0.75 &     0.39 & e &  0.71 &  0.86 & & \\
 &   2.3 &  -5.6 &  6.1/1.1 &    23.83  &    22.51  &   -0.69 &     0.50 &     0.82 & e &  0.51 &  0.62 & & \\
 &  -1.1 &  -6.4 &  6.5/1.2 &    23.67  &    22.08  &   -0.24 &     0.39 &     1.20 & e &  0.47 &  0.88 & & \\
 &  -8.5 &   2.8 &  9.0/1.7 &    23.18  &    22.84  &   -0.65 &     0.03 &     0.31 & e &  0.39 &  0.28 & & \\
 &  11.4 &  -6.3 & 13.1/2.5 &    23.56  &    22.48  &   -0.78 &     0.37 &     0.71 & e &  0.05 &  0.09 & & \\
 &   0.2 & -14.7 & 14.7/2.8 & $>$25.00  &    23.22  &         &          & $>$ 1.08 & e &       &  0.01 & & \\
 &   0.0 &  16.2 & 16.2/3.0 &    23.51  &    22.60  &   -0.80 &     0.24 &     0.67 & e &  0.01 &  0.01 & & \\
 &  -4.8 & -17.6 & 18.2/3.4 &    23.10  &    21.46  &   -0.34 &     0.77 &     0.87 & e &  0.00 &  0.01 & & \\
 &  -6.3 &  18.0 & 19.0/3.6 &    24.22  & $>$23.50  &   -0.72 &     0.14 & $<$ 0.58 & e &  0.00 &       & & \\
  &  17.3 &   8.8 & 19.4/3.6 &    18.93  & $<$18.02  &   -0.15 &     0.55 & $>$ 0.36 & p &  0.01 &  0.00 & 0.00  & G~star\\
\hline    
 X306--42 &   3.1 &  -3.6 &  4.7/1.0 &    24.08  & $>$23.50  &   -0.74 &     0.11 & $<$ 0.47 & e &  0.44 &       & & \\
 &  -3.6 &   5.6 &  6.6/1.5 &    22.62  &    22.04  &   -0.60 &     0.34 &     0.24 & p &  3.18 &  1.23 & 1.065 & {\bf AGN}\\
 &  -8.0 &   2.1 &  8.2/1.8 &    24.03  &    23.29  &   -0.25 &     0.02 &     0.72 & e &  0.15 &  0.15 & & \\
 &  10.4 &  -5.4 & 11.7/2.6 &    23.37  &    22.47  &   -0.60 &     0.36 &     0.54 & e &  0.08 &  0.09 & & \\
 &  -2.0 &  12.2 & 12.3/2.7 & $>$25.00  &    23.40  &         & $>$ 0.34 &     1.26 & p &       &  0.02 & & \\
 &  10.4 &  11.8 & 15.7/3.4 &    22.03  &    21.19  &   -0.10 &     0.54 &     0.30 & e &  0.01 &  0.01 & 0.078 & NELG\\
\hline    
 X051--43 &   1.1 &  -1.1 &  1.5/0.2 &    21.44  &    20.60  &   -0.12 &     0.48 &     0.36 & p &  9.00 &  5.70 & ?.??? & {\bf Bl~Lac}~(?)\\
 &  -6.3 &   6.9 &  9.4/1.4 &    24.45  &    23.55  &$>$-0.95 & $<$ 0.15 & $>$ 0.75 & e &  0.13 &  0.10 & & \\
 &  -9.7 &   5.2 & 11.0/1.7 &    23.39  & $>$23.50  &   -0.50 &    -0.05 & $<$-0.06 & e &  0.27 &       & & \\
 &  -0.6 &  13.9 & 13.9/2.1 &    24.28  &    22.68  &$>$-0.78 &     0.78 &     0.82 & e &  0.04 &  0.09 & & \\
 & -13.1 &   4.8 & 14.0/2.1 &    23.71  &    23.59  &$>$-0.21 &    -0.41 &     0.53 & e &  0.07 &  0.03 & & \\
 &   3.8 & -17.4 & 17.8/2.7 &    24.22  & $>$23.50  &$>$-0.72 & $<$-0.08 &          & e &  0.01 &       & & \\
 & -12.2 & -14.5 & 19.0/2.9 &    20.23  &    19.20  &   -0.13 &     0.60 &     0.43 & e &  0.04 &  0.01 & 0.096 & Starburst\\
 & -11.7 &  16.6 & 20.3/3.1 &    23.86  &    23.51  &$>$-0.36 &    -0.30 &     0.65 & e &  0.01 &  0.00 & & \\
 &  -6.1 &  21.2 & 22.0/3.3 &    23.00  &    21.52  &$>$ 0.50 &     0.81 &     0.67 & p &  0.01 &  0.01 & & \\
\hline    
 X407--44 &  -3.3 &  -4.7 &  5.7/0.8 &    21.41* &    20.65  &   -0.94 &     0.19 &     0.57 & p &  5.68 &  3.59 & 1.821 & {\bf AGN$^{(1)}$}\\
 &  14.1 &   0.9 & 14.2/2.0 & $>$23.50* &    21.08  &         &          & $>$ 1.42 & p &       &  0.67 & & \\
 & -18.2 & -16.0 & 24.2/3.3 &    22.73* &    20.99  &   -0.92 &     0.93 &     0.81 & e &  0.00 &  0.00 & & \\
\hline    
 X233-45 &   4.9 &  -5.8 &  7.6/1.1 &    24.86  &    23.72  &   -1.30 & $<$ 0.56 & $>$ 0.58 & e &  0.09 &  0.11 & & \\
 &  -8.4 &  -2.5 &  8.8/1.2 &    24.83  &    23.33  &   -1.29 &     0.71 &     0.79 & e &  0.07 &  0.14 & & \\
 &  -4.0 &  -8.8 &  9.6/1.3 &    22.55  &    21.86  &   -0.82 &     0.19 &     0.50 & e &  0.29 &  0.48 & ?.??? & low S/N\\
 &  -9.2 &   5.9 & 10.9/1.5 &    24.69  &    21.76  &$>$-1.19 &     1.64 &     1.29 & e &  0.06 &  0.37 & & \\
 & -11.0 &   3.2 & 11.5/1.6 &    24.86  &    23.10  &$>$-1.36 &     0.57 &     1.19 & e &  0.04 &  0.08 & & \\
 & -13.3 &  -1.8 & 13.4/1.9 &    22.51  &    21.70  &   -0.66 &     0.25 &     0.56 & e &  0.12 &  0.20 & 1.180 & {\bf AGN}\\
 & -12.0 & -11.9 & 16.9/2.4 &    21.29  &    18.62  &    1.19 &     1.51 &     1.16 & e &  0.49 &  0.21 & & \\
 & -16.1 &  10.4 & 19.2/2.7 &    24.49  & $>$23.50  &   -0.49 & $<$ 0.19 &          & e &  0.01 &       & & \\
 &   6.3 &  20.7 & 21.6/3.0 &    22.41  &    21.12  &   -0.54 &     0.58 &     0.71 & e &  0.01 &  0.01 & & \\
 & -17.6 & -17.0 & 24.5/3.4 &    24.50  &    23.34  &   -0.69 & $<$ 0.20 & $>$ 0.96 & e &  0.00 &  0.00 & & \\
\hline
 X109-46 &  -1.9 &  -0.2 &  1.9/0.4 &    23.56  &    22.59  &   -0.52 &     0.42 &     0.55 & p &  2.25 &  0.58 & & \\
 &   2.6 &   3.6 &  4.4/0.9 &    23.63  &    23.10  &   -0.52 &     0.29 &     0.24 & e &  0.80 &  0.44 & & \\
 &  -6.2 &   5.1 &  8.0/1.7 &    21.70  &    20.56  &   -0.54 &     0.75 &     0.39 & e &  0.74 &  0.60 & 0.269 & Starburst\hfill(C)\\
 &   2.5 &   7.6 &  8.0/1.7 &    24.60  &    22.96  &$>$-1.10 &     1.04 &     0.60 & e &  0.11 &  0.36 & & \\
 &   2.1 &  -9.2 &  9.4/2.0 &    24.30  &    21.36  &$>$-0.80 &     1.44 &     1.50 & p &  0.10 &  1.51 & 0.00  & M~star\hfill(E)\\
 &   4.0 &  -9.8 & 10.6/2.2 &    22.79  &    20.06  &$>$ 0.71 &     1.57 &     1.16 & p &  0.71 &  0.79 & 0.00  & M~star\hfill(F)\\
 &   7.5 &   9.1 & 11.8/2.5 &    24.78  & $>$23.50  &$>$-1.28 & $<$ 0.48 &          & e &  0.02 &       & & \\
 &   0.5 & -12.7 & 12.7/2.7 &    24.75  &    23.44  &$>$-1.25 &     0.83 &     0.48 & e &  0.01 &  0.02 & & \\
 &  -4.9 &  14.7 & 15.5/3.2 &    22.78  &    21.83  &   -0.83 &     0.25 &     0.70 & e &  0.01 &  0.01 & 0.626 & NELG\hfill(I)\\
\hline    
 X213--47 &  -8.5 & -13.4 & 15.8/2.9 &    24.70  & $>$23.50  &$>$-1.20 & $<$ 0.40 &          & e &  0.00 &       & & \\
 & -13.8 & -13.8 & 19.5/3.5 &    24.51  &    23.54  &$>$-1.01 &     0.25 &     0.72 & e &  0.00 &  0.00 & & \\
\hline    
\end{tabular}
\end{table*}
\begin{table*}     
\begin{tabular}{rrrrrrrrrcrrll}
\hline
\\
 ID.  & $\Delta\alpha$ & $\Delta\delta$ & $\Delta$\arcs/$\displaystyle{\Delta\over\varepsilon_{\rm{x}}}$ &
 \multicolumn{1}{c}{~m$_B$} & \multicolumn{1}{c}{m$_R$} & U-B & B-V & V-R & &
 \multicolumn{1}{c}{$LR_B$} &\multicolumn{1}{c}{$LR_R$} & \multicolumn{1}{c}{z} & object notes\\
 &\multicolumn{1}{c}{~~[\arcs]}  & \multicolumn{1}{c}{~~[\arcs]} \\
\hline     
 X022--48 &  -1.7 &  -1.3 &  2.1/0.5 &    22.21  &    19.86  &    0.72 &     1.40 &     0.95 & e &  3.25 &  3.36 & 0.32: & {\bf Interactive~gal}\\
 &  -1.6 &  -4.6 &  4.9/1.1 &    23.31  &    22.06  &   -0.20 &     0.74 &     0.51 & e &  1.30 &  1.47 &       & companion\\
 &   5.9 &  -6.0 &  8.5/2.0 &    24.04  & $>$23.50  &$>$-0.54 &    -0.03 & $<$ 0.57 & e &  0.12 &       & & \\
 &   8.9 &  -7.9 & 11.9/2.7 &    22.33  &    21.00  &   -0.38 &     0.66 &     0.67 & e &  0.09 &  0.07 & 0.474 & NELG\\
 &  -3.3 &  11.5 & 12.0/2.8 &    21.90  &    20.70  &   -0.38 &     0.64 &     0.56 & e &  0.08 &  0.07 & 0.389 & early+[OII]\\
 & -10.0 &  -9.6 & 13.9/3.2 &    23.76  &    22.55  &   -0.25 &     0.71 &     0.50 & e &  0.01 &  0.01 & & \\
 & -15.4 &  -0.6 & 15.4/3.5 &    24.51  & $>$23.50  &$>$-1.01 & $<$ 0.21 &          & e &  0.00 &       & & \\
 &  -6.5 & -14.9 & 16.3/3.7 &    21.03  &    20.01  &    0.01 &     0.64 &     0.38 & p &  0.02 &  0.01 & 0.00  & F/G star\\
\hline    
 X215--49 &   5.4 &  -7.4 &  9.2/1.7 &    23.12  &    22.44  &   -0.69 &     0.18 &     0.50 & e &  0.37 &  0.42 & & \\
 &  -0.1 & -14.3 & 14.3/2.7 & $>$25.00  &    23.32  &         &          & $>$ 0.98 & e &       &  0.01 & & \\
 &   3.7 &  14.9 & 15.4/2.9 &    22.58  &    22.19  &   -0.78 &     0.19 &     0.20 & e &  0.02 &  0.03 & 1.053 & strong\hfill[OII] gal~(C)\\
 &  -3.1 &  16.4 & 16.6/3.1 &    23.62  &    22.47  &   -0.60 &     0.59 &     0.56 & e &  0.01 &  0.01 & & \\
 &  17.6 &   0.0 & 17.6/3.3 &    24.35  &    22.93  &$>$-0.85 &     0.77 &     0.65 & e &  0.00 &  0.00 & & \\
\hline    
 X264--50 &   0.4 &   3.5 &  3.5/0.7 &    24.44  & $>$23.50  &$>$-0.94 & $<$ 0.14 &          & e &  0.45 &       & & \\
 &  -2.1 &  -7.2 &  7.5/1.4 &    21.84  &    20.77  &   -0.42 &     0.49 &     0.58 & e &  0.89 &  0.73 & 0.568 & {\bf Starburst}\\
 &   3.1 &  -9.1 &  9.6/1.8 &    23.46  &    22.20  &$>$ 0.04 &     0.66 &     0.60 & e &  0.31 &  0.36 & & \\
 &  -7.6 &  -7.5 & 10.7/2.0 &    24.60  & $>$23.50  &$>$-1.10 &     0.10 & $<$ 1.00 & e &  0.05 &       & & \\
 &  -9.9 &  -8.7 & 13.2/2.5 &    23.24  &    22.96  &   -0.41 &    -0.03 &     0.31 & p &  0.14 &  0.02 & & \\
 &  17.8 &   2.1 & 17.9/3.4 &    23.53  &    22.06  &   -0.17 &     0.64 &     0.83 & e &  0.00 &  0.01 & & \\
\hline
\end{tabular}
\end{table*}

The LR values for most of these 15 suggested identifications are greater
than 1 and are the highest in their error boxes. The only exceptions are 
the broad--line radio--galaxy identified with the source X409--28 and the AGN 
identified with the source X032--36, whose LR values are of the order of 0.5. 
In the latter error box there is also a brighter stellar object with 
higher LR values. However, its spectrum shows that it is a B~type star 
and its identification with the X-ray source can therefore be excluded 
on the basis of the $f_x/f_v$ ratio (Maccacaro et~al. 1988). 
A few other objects have been observed spectroscopically in these 15
error boxes (see last column in Table 3) and most of them have been found to be
narrow emission line galaxies (NELG) or starburst galaxies. In all cases
their LR value is significantly smaller than that of the best candidate and
therefore we consider them unlikely to be associated to the
X-ray sources. The estimated surface densities of AGNs with 
 $ m_B \le$ 23.5 ($\sim$ 325 per square degree, Mignoli and Zamorani 1998) 
is similar to the surface density of M stars with $m_V < $ 20. From the
sum of the two surface densities we estimate that only
$\sim$ 0.7 random coincidences in the
95\% error circles of the 31 X-ray sources are expected for AGNs and M stars
in the magnitude ranges covered by these suggested identifications.

 iii. For the 16 error boxes which at this stage were still without
a reliable optical identification, we have taken spectra also of objects
classified as extended. At the typical magnitudes of the optical
counterparts ($m_B >$ 22.0) the surface density of objects classified as
extended in our optical images is higher than that of stellar objects. The 
ratio between the numbers of extended and stellar objects in our CCD
catalogue is $\sim$ 2 at $m_B \sim$ 22.25 and $\sim$ 5 at $m_B \sim$ 23.25. 
Therefore, the large number of extended objects
makes more difficult to deal with the problem of random coincidences and to find
convincing optical identifications. Each of these sixteen error boxes is now
discussed in some detail, in order of decreasing X-ray flux. 

{\bf X043--12}: the 95\% error box contains only one object, classified
as extended in the blue band and as point-like in the V and R bands. Its noisy
spectrum shows two narrow lines well coincident with Ly$\alpha$ and CIV at z 
= 2.80. It appears to be a high redshift analogue of the low--z type--2 
AGNs, similar to the QSO~2 at z = 2.35 found by Almaini et al. (1995)
in the optical follow--up of other deep ROSAT fields. Although a higher quality
spectrum would be needed to confirm the nature of this object, given its
relatively large LR values, we consider it as the correct identification.

{\bf X039--19} and {\bf X049--21}: these are the brightest X-ray sources in our sample without any
suggested spectroscopic identification. The 95\% error box of X039--19 does not 
contain any object in our CCD data, while a faint ($m_B \sim 24$), extended
object is present near the center of the error box of X049--21. Its relatively
large LR values suggest that it might be the correct optical identification. 
However, both the spectrum of this object (shown in figure 4) and the
spectrum of a similarly faint object just outside the error box of X039--19 
(object A in figure 4) have extremely low S/N and do not allow any 
spectroscopic classification. Moreover, both X-ray sources have a close--by
source at less than one arcmin (see figure 1), so that it is possible that 
their X-ray positions are not as well determined as those of the other 
sources with similar X-ray flux.

{\bf X211--22}: we took spectra for three objects, all of them just outside
the 95\% radius. The brightest one is an early type
galaxy at z = 0.180, while the other two, separated by about 2 arcsec
from each other, 
are a Sey 1 galaxy (classified as point--like in the B band and as extended
in the R band) at z = 0.281 and a B star. We consider the Sey 1 galaxy as the
most likely identification.

{\bf X404--23}: its large error box makes the identification difficult. The 
3$\sigma$  error box contains 18 sources and all
of them have low LR values. We took spectra for three
objects, finding a NELG galaxy (object G) and a Sey 2 galaxy (object N)
at different redshifts, while the third one (object J) is a faint 
($m_B = 23.74$) M star. None of them is a 
convincing identification. Visual inspection of the X-ray image suggests
the presence of two different sources, at about 40\arcs from each other,
which our detection algorithm has not been able to separate. This is also
supported by the large positional error resulting from the maximum likelihood
fit. Since the two sources appear to have approximately the same number
of counts, it is likely that both of them are very close to or just
below our X-ray detection threshold.
\begin{figure*}[p]
\epsfysize=24.5cm
\vspace{-1.0cm}
\vspace{22.cm}
\caption[]{Finding charts for the X-ray sources with 68\% and 95\% error boxes
and spectra of the most likely identifications.} 
\end{figure*}
\begin{figure*}[p]
\epsfysize=24.5cm
\vspace{-1.0cm}
\vspace{22.cm}
\addtocounter{figure}{-1}
\caption[]{continued}
\end{figure*}
\begin{figure*}[p]
\epsfysize=24.5cm
\vspace{-1.0cm}
\vspace{22.cm}
\addtocounter{figure}{-1}
\caption[]{continued}
\end{figure*}
\begin{figure*}[p]
\epsfysize=24.5cm
\vspace{-1.0cm}
\vspace{22.cm}
\addtocounter{figure}{-1}
\caption[]{continued}
\end{figure*}
\begin{figure*}[p]
\epsfysize=24.5cm
\vspace{-1.0cm}
\vspace{22.cm}
\addtocounter{figure}{-1}
\caption[]{continued}
\end{figure*}
\begin{figure*}
\epsfysize=24.5cm
\vspace{-1.0cm}
\vspace{22.cm}
\addtocounter{figure}{-1}
\caption[]{continued}
\end{figure*}

{\bf X408--30}: the radio source coinciding with X408-30 and indicated with
a cross in figure 4 is not associated to any optical object 
(see Gruppioni, Mignoli and Zamorani 1999). If the radio and X-ray sources 
are really associated, then the optical counterpart of the X-ray source
is fainter than our optical limit. The brightest object in
the field (object C), but well outside the 95\% error box, is a 
NELG at z = 0.814.

{\bf X207--31}: all the objects in the error box are extended. We took
spectra for two galaxies, one ($m_R$ = 22.43) just outside the 1$\sigma$ error
box and the second one ($m_R$ = 19.65)  just outside the 3$\sigma$ error box. 
Both of them have z $\sim$ 0.58; the second galaxy, which is the brightest
and reddest of all the surrounding galaxies appears to be a cD in a cluster.
(Figure 4 shows the spectrum of this galaxy).
Just below our maximum likelihood threshold for X-ray detection, at about 
two arcmin from X207--31, there is an other X-ray source in a position where an
overdensity of faint galaxies is clearly seen. Three of these galaxies
have the same redshift as those in X207--31. We therefore identify X207--31
with a cluster, probably interacting with a second cluster at a distance of 
about 1.5 Mpc.

{\bf X250--34}: the error box contains only faint ($m_B \ge$ 23.2,
$m_R \ge$ 22.9) extended objects, none of which with a large LR. No spectrum
was taken for any object.

{\bf X011--35}: the density of objects in the error box is twice as
high as the average density in our catalog. We took spectra for the
six brightest objects ($m_R <$ 21.5) finding three galaxies at z $\sim$ 0.39 
(objects D, L  and O in figure 4), one absorption line galaxy at z = 0.189 
(object A; this is the object closest to the X-ray position), 
one early--type galaxy at z = 0.586 (object F). 
The redshift of this galaxy is about the same as that of the
clusters discussed in connection with X207--31. The angular distance between
X011--35 and X207--31 ($\sim$5 arcmin) corresponds to about 4 Mpc, suggesting
the existence of a large scale structure at this redshift. The sixth 
spectrum (object I) has a very low S/N and no redshift was derived. 
The spectrum in figure 4 is the spectrum of the galaxy closest to the 
X--position among those at z $\sim$ 0.39 (D). On the basis of these results, 
although with some possible ambiguity, we identify the X-ray source with 
a group of galaxies. 

{\bf X234--41}: also in this error box the density of faint extended
objects is twice as high as the average density. In particular, 
in the inner 10\arcs there are six extended objects with $ 23 \le m_B \le 24 $,
while about one would be expected. 
Although
no spectrum was taken, because of the faintness of the objects, we tentatively
identify this source with a group or cluster of galaxies. The bright object
just outside the 3$\sigma$ error box is a G star with a too large X-ray to
optical ratio to be associated with the X-ray source.

{\bf X233--45}: we took spectra of the two brightest objects
($m_B \sim$ 22.5) within 15\arcs from the X-ray position. Both of them are 
classified as extended. One spectrum is very blue, with low S/N and no redshift
was determined; the second object is an AGN at z = 1.180, which we consider
to be the identification. No spectrum is available for the more distant
and brighter extended object, whose colours suggest a low redshift
elliptical galaxy.

{\bf X109--46}: the error box contains three relatively bright
objects ($m_R <$ 21.5), two of which are point--like and one extended. The
two stellar objects are two M stars (objects E and F in figure 4), 
while the extended one is a starburst galaxy at z = 0.269 (object C). 
The two M stars appear to be too faint for being the counterpart of the 
X-ray source. The starburst galaxy might be considered a likely candidate. 
We note, however, that very close to the X-ray position, there are two faint,
blue objects with LR values larger than that of the starburst galaxy. One of 
these two objects, classified as point--like, is a good candidate
for being an AGN. Waiting for spectra for these objects, we consider
this source not identified yet.

{\bf X213--47}: no object appears in our optical catalogs within the
95\% error box, although a very faint object ($m_R \ge$ 24.0) is barely
detected in two different exposures of the R CCDs at $\sim$ 2 arcsec from
the X-ray position.

{\bf X022--48}: the noisy spectrum of the galaxy close to the center
of the X-ray error box shows a narrow emission line which, if interpreted 
as [OII], corresponds to z = 0.32. Both the blue and the red CCDs show clear 
signs of interaction with a fainter, blue extended object located about 3\arcs
south. Also on the basis of the LR values, we consider this complex of 
interacting galaxies as a likely identification. The two galaxies just outside
the 2$\sigma$ error box have different redshifts, while the brighter, more
distant point--like object is an F/G star.

{\bf X215--49}: no obvious candidate is contained within the 95\% error
box. The spectrum of the galaxy just outside of it (object C)
shows a strong [OII] emission line and MgII2800 absorption line at z = 1.053.
Its LR is small and we do not consider this object as the correct
identification.

{\bf X264--50}: there is one object in the error box with LR $\sim$ 1.
 It is a relatively bright ($m_R$ = 20.77) starburst galaxy at z = 0.568.
Despite the presence of a much fainter
($m_R \sim $ 23) blue, point-like source which may be an AGN candidate,
we consider the galaxy as a possible identification.

\section{ Discussion }
\subsection {Summary of identifications and not identified sources}

On the basis of the criteria discussed above, we have found reliable 
spectroscopic identifications for 41 sources (82\% of the total),
indicated in bold face in the last column of Table 3.
For one more source (X234--41) the identification is very likely to be with 
a faint cluster of galaxy, although no redshift is available. These 42 
reliable identifications are 33 AGNs (including the two radio galaxies and
the BL Lac candidate; 79\% of the identified sources), 2 galaxies, 3 groups or
clusters of galaxies and 4 stars. Except for the higher fraction of unidentified
sources (see discussion below), the identification content of this sample
is in excellent agreement with what has been found, at a similar flux
limit, in the much deeper PSPC and HRI surveys in the Lockman field (Hasinger
et al. 1998, Schmidt et al. 1998). 

Figure 5 shows the expected and
observed cumulative distributions of the distances normalized to the error 
on each coordinate between the optical counterparts and the X-ray sources.
In the observed distribution we have excluded the three sources
identified with groups or clusters of galaxies, because for these objects
it is more difficult to unambiguously define the optical position. The excellent
agreement between the two distributions shows the goodness of our 
derived ML estimates for the positional uncertainties, at least when the
X-ray position is not affected by confusion problems (see below).  
\begin{figure}[t]
\epsfysize=9cm
\epsfbox{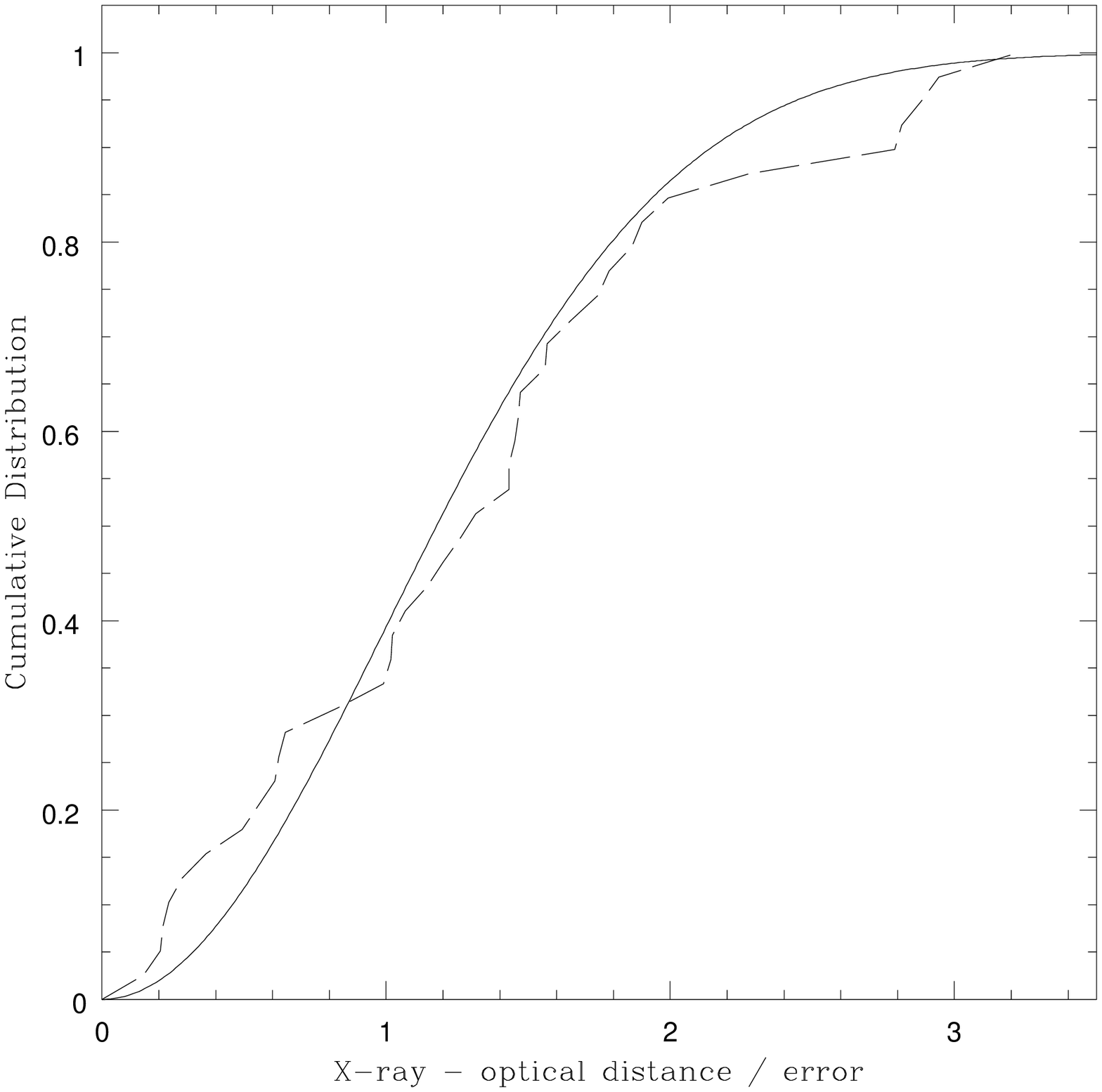}
\caption[]{ Cumulative observed distribution (dashed curve) of the distances 
normalized to the error on each coordinate between the optical counterparts 
and the X-ray sources. Here we have excluded the three sources identified
with groups or clusters of galaxies. The solid curve shows the expected
distribution.}
\end{figure}

If we divide our sample into two equally populated sub--samples as a 
function of flux (S$> 6.5 \times 10^{-15}$ and 
S$< 6.5 \times 10^{-15}$ \erg), we find that
the percentage of identifications remains approximately constant (88\% and
80\% in the high and low flux sub--samples, respectively). 
AGNs are the dominant class of objects in both sub--samples (90\% of the 
optical identifications in the high flux sub--sample and 65\% in the low flux
sub--sample), while the few identifications with clusters and galaxies are
all in the low flux sub--sample. With the two extreme assumptions that none 
or all of the still unidentified objects are AGNs we derive that the 
percentage of AGNs among {\bf all} the X-ray sources in our sample 
is comprised between 66\% and 82\%. 
\begin{figure}[t]
\epsfysize=9cm
\epsfbox{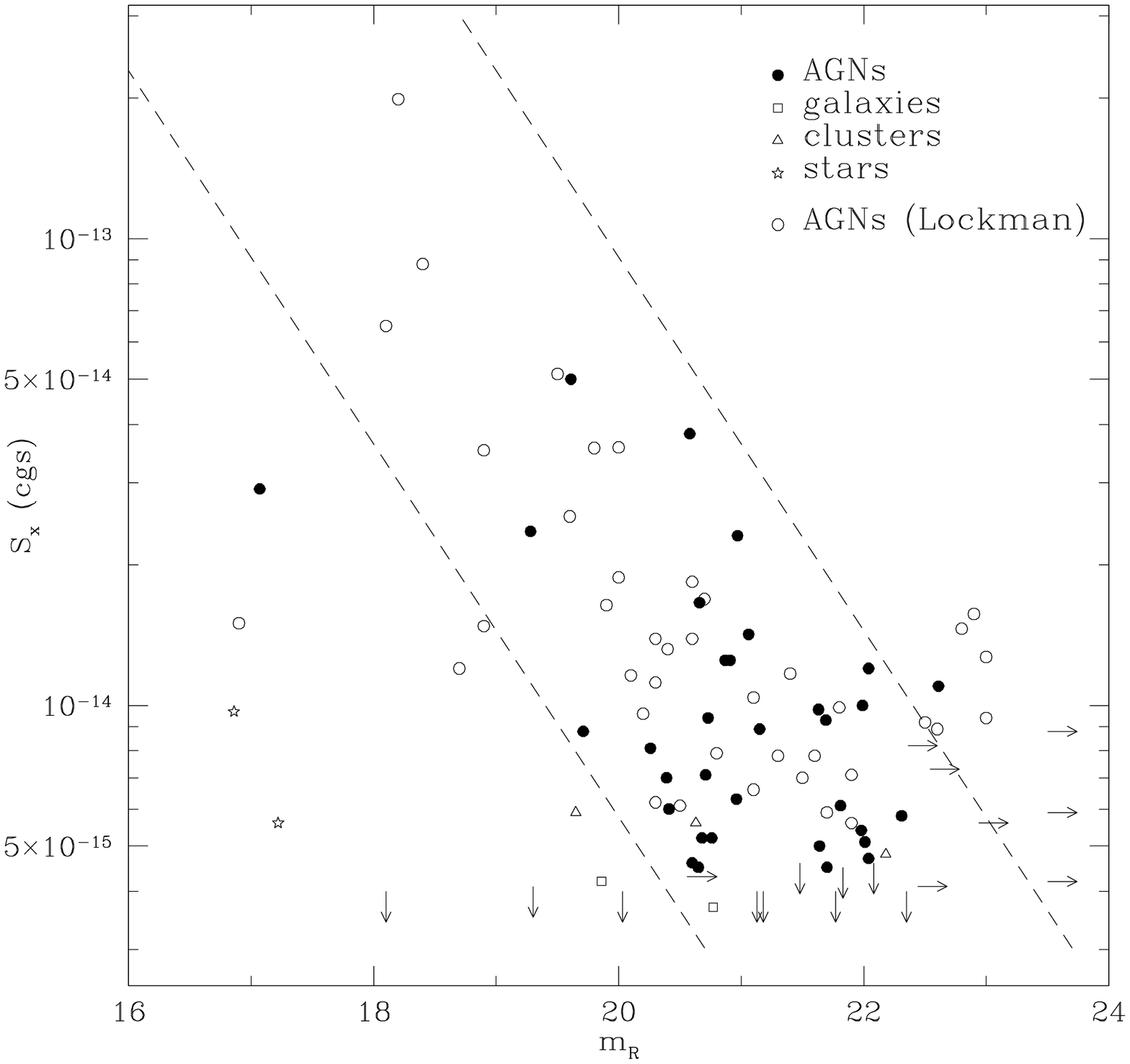}
\caption[]{ X-ray flux versus $m_R$ magnitudes for all the sources in our
sample and for the AGNs identified in the
Lockman Field (Schmidt et al. 1998). The two straight dashed lines,
corresponding to constant X-ray to optical ratios, show the locus of this
plane (-0.6 $< log~f_x/f_v < $ 0.6) which contains most of the identified 
AGNs in both samples. For a description of the upper limits, both in optical
and in X-ray, see text.}
\end{figure}

Figure 6 shows the X-ray flux versus $m_R$ magnitudes for the sources in our
sample and for the AGNs identified in the Lockman Field (Schmidt et al. 1998). 
The two bright F stars ($m_B \sim$ 10) are not shown here. 
The two straight dashed lines, corresponding to constant X-ray to optical 
ratios, show the approximate locus of this plane which contains most of the 
identified AGNs, both in our sample and in the Lockman Field.
This range in X-ray to optical fluxes correspond to 
-0.6 $< log~f_x/f_v < $ 0.6, where  $ log~f_x/f_v $ is defined as in 
Maccacaro et al. (1988). For any given X-ray flux the range in magnitude 
is about $\pm$ 1.5, corresponding to about $\pm$ a factor of 4
with respect to the average X-ray to optical ratio. Despite the much higher 
X-ray flux limit and, correspondingly, much brighter limiting magnitudes, 
also most ($\sim$ 90\%) of the EMSS AGNs lie in the same band.
For the three sources identified with groups or clusters of galaxies we
have plotted the magnitude of the brightest galaxies. The two X-ray
sources identified with galaxies have an X-ray to optical ratio close to the
lower limit of those of AGNs. 

The limits for the eight unidentified sources
are plotted at the magnitude of the brightest object within the 95\% error
box which is not excluded from being the optical counterpart on the
basis of the available spectra. Three of these upper limits lie outside 
the band shown in the figure and three more are very close to the upper
bound of this band. If associated with AGNs, they would have an X-ray to 
optical ratio significantly higher than the average. As shown in the figure, 
6 such AGNs have been identified in the Lockman field; four of them are 
classical broad line objects, while the other two, showing only narrow lines, 
have been classified as AGNs on the basis of the presence of [NeV]$\lambda$3426
emission (Schmidt et al. 1998). The HRI arcsec positions in the Lockman field
has made these identifications with such faint optical objects possible.
In our case, with the 95\% error radius for these sources ranging from
8\arcs to 24\arcs, similar identifications are significantly more difficult.
Alternatively, we can not exclude that some of these sources may have
a ``wrong'' position because of confusion. In Section 2 we have estimated
that in about 2--5 cases a single source in our list may be produced by two 
different close--by sources, so that its position may be significantly wrong 
and we may not be able to find any reliable optical counterpart within the 
error box. The much more detailed simulations performed by 
Hasinger et al. (1998) show that in a PSPC survey like ours up to 
almost 20\% of the detected sources with $S_x < 1 \times 10^{-14}$ \erg 
(corresponding to about 7 sources) may appear at a detected position more 
than 15\arcs away from the true position because of confusion. From these 
considerations it follows that only higher resolution, deep X-ray data 
(e.g. with AXAF and XMM) can fully clarify the situation for these sources.
\begin{figure}[t]
\epsfysize=9cm
\epsfbox{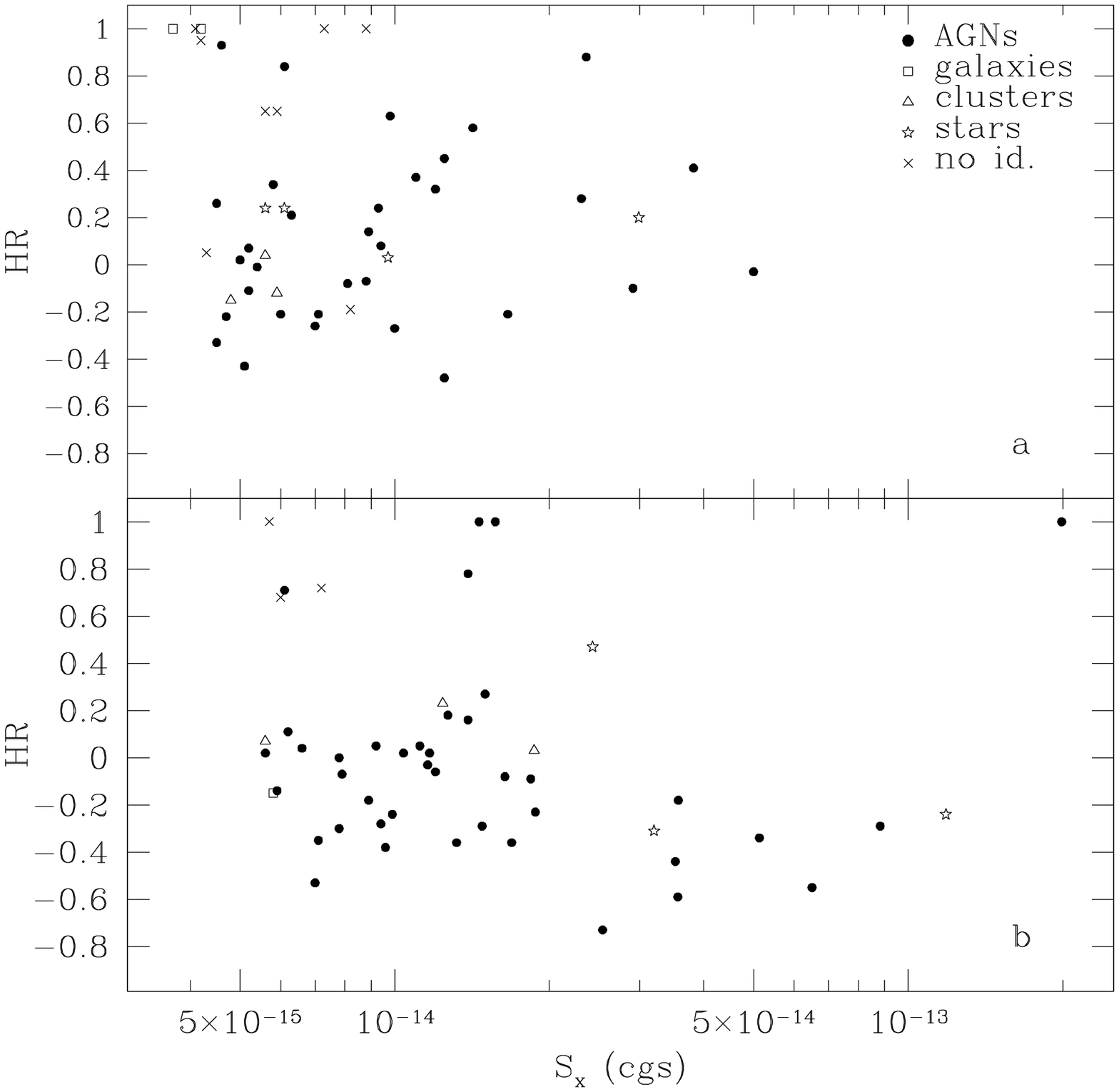}
\caption[]{ Hardness ratio versus X-ray flux for the sources in the Marano
field (panel a) and in the Lockman field (panel b). The symbols for the various
classes of objects are shown in panel a.
}
\end{figure}

Some additional clues on the nature of the unidentified sources can be
obtained from the analysis of the hardness ratio. Figure 7 shows the
hardness ratio versus X-ray flux for the sources in the Marano
field (panel a) and in the Lockman field (panel b). 

Analysis of this figure shows the following results:

i. Most of the sources in both fields occupy well defined bands in HR. On
average, the observed hardness ratio values in the Lockman field are smaller 
than those in the Marano field. The reason for this shift in HR is 
essentially due to the lower $N_H$ value in the Lockman field, so that the
observed spectra appear softer. When converted to energy spectral indices, 
under the assumption of a single power law with the galactic $N_H$, 
the HR band in the Marano field corresponds to the range 
0.5 $\le \alpha_x \le $ 2.0. Approximately the same range in $\alpha$
is derived for the HR band in the Lockman field assuming $N_H \sim 1 \times
10^{20}$, i.e. approximately twice the value of atomic hydrogen
derived from measurements in the 21 cm line (Lockman et al. 1986). 
A similar $N_H$ value was derived
by Hasinger et al. (1993) when fitting the average spectrum
of all the X-ray sources in the Lockman field. This excess
of X-ray absorption is consistent with the column density of ionized gas
in the galactic disk derived from analyses of the dispersion measures
of pulsars (Reynolds 1989). The presence of this additional column of
ionised gas should be taken into account when analyzing ROSAT data,
particularly in directions of low $N_H$ as, for example, in the Lockman field.

ii. In both samples there are about (15--20)\% of the sources which appear
to have hard or absorbed spectra, with HR values close to 1. These sources
appear to be reasonably well separated from the others, especially in the
Lockman field where, because of the higher S/N ratio, the errors in HR are
smaller. The fraction of sources without optical identification is significantly
higher among these hard sources than among sources with ``normal'' X-ray
spectrum. For example, in the Lockman field the unidentified sources
 are 3 out of 8 for sources
with HR $>$ 0.6, to be compared with zero out of 42 for sources with
HR $<$ 0.6; in the Marano field the corresponding fractions of unidentified 
sources in the same HR ranges are 6/12 and 2/38. 
The eleven identifications (Lockman and
Marano fields together) among these hard sources comprise 2 galaxies
(X264--50 and X022--48), and 9 AGNs (4 broad line AGNs, 2 radio galaxies, 1
possible BL Lac, and 2 narrow line objects classified as AGNs on the basis of
the presence of [NeV]$\lambda$3426). This shows that AGNs
are the dominant population also in this range of HR. However, contrary
to what happens for objects with ``normal'' X-ray spectra, a relatively large
fraction of them does not show the prominent broad lines typical of
classical quasars. The absence of broad lines coupled with the
presence of X-ray absorption are consistent with the standard unified models
for AGNs. It is interesting to note that the two narrow line objects 
with [NeV]$\lambda$3426 have two of the highest X-ray to optical ratios
in the Lockman field sample. Since the current optical limits for the
unidentified sources in the Marano field correspond to X-ray to optical ratios 
similar to those of these sources, it is possible that at least some of our
hard, unidentified sources belong to the same class of objects.

iii. In both samples, but especially in the Marano field, the hard sources tend 
to be more numerous at low flux levels. While this is likely to be, at least
in part, a real effect, as shown by the identification content of these 
sources discussed above, there is also an obvious selection effect
which favours hard spectra near the detection limit. At these fluxes, because
of the combined effect of confusion and statistical fluctuations on small
number of counts, a not--negligible fraction of the detected sources has
measured fluxes significantly higher than the true fluxes (see Hasinger et al.
1998). Since the sample is defined in the ROSAT hard band, the measured
hardness ratio for some of these sources would be biased toward large
values and therefore the observed fraction of hard sources is probably
higher than the real one.

\subsection{Optically and X-ray selected AGNs}

As mentioned in Section 1, in the same area covered by the X-ray survey we 
have conducted in the past years a search for optically selected AGNs, using
multi-colour data from plates taken at the ESO 3.6m telescope. 
(Zitelli et al. 1992). This survey has later been extended to fainter
magnitudes using CCD data (Mignoli et al. in preparation) and has produced,
so far, 
spectroscopic data for 29 optically selected broad--line AGNs inside
the ROSAT area. Ten of these have not been detected in the X-ray data and
their corresponding X-ray upper limits are shown in Figure 6. Three of these
AGNs have X-ray upper limits outside the band shown in Figure 6,
in the area of very low X-ray to optical ratio. 

On the basis of the available CCD and spectroscopic data, which do not cover
the entire ROSAT area of about 0.2 sq.deg., Mignoli and Zamorani (1998)
estimate surface densities of $\sim$ 185 and 140 AGNs per sq.deg. with
$m_B < 22.5$ and $22.5 < m_B < 23.5$, respectively. These estimates
are in good agreement with the predictions obtained by Zamorani (1995)
on the basis of reasonable extrapolations from counts at slightly
brighter magnitudes. Limiting ourselves at the classical broad--line AGNs
and merging together the X-ray and the optically selected samples, we 
have a total of 35 such objects with $m_B \le$ 22.6 corresponding to a
surface density of 178 $\pm$ 30 broad--line AGNs per sq.deg. 
This is the highest reported surface density for these objects so far 
at this magnitude.  
Taking into account that, as mentioned above, the optically selected
sample is not complete over the entire ROSAT area, this density
is significantly higher than the estimate of about 125 AGNs
per sq.deg. with $m_B \le 22.6$ recently obtained in the Deep Multicolor Survey,
on the basis of 53 spectroscopically confirmed AGNs (Kennefick et al.
1997). Our estimated surface densities for $m_B < 22.5$ and 
$22.5 < m_B < 23.5$, correspond to $36 \pm 6$ and $27 \pm 5$ AGNs inside the 
ROSAT area. Since in the same magnitude ranges ROSAT has detected 23
and 6 broad--line AGNs respectively,
we conclude that the ``efficiency'' of AGN selection with X-ray
exposures reaching about $4 \times 10^{-15}$ \erg is
$\sim$ 65\% and $\sim$ 20\% in the two magnitude ranges.
 
\begin{figure}[t]
\epsfysize=9cm
\epsfbox{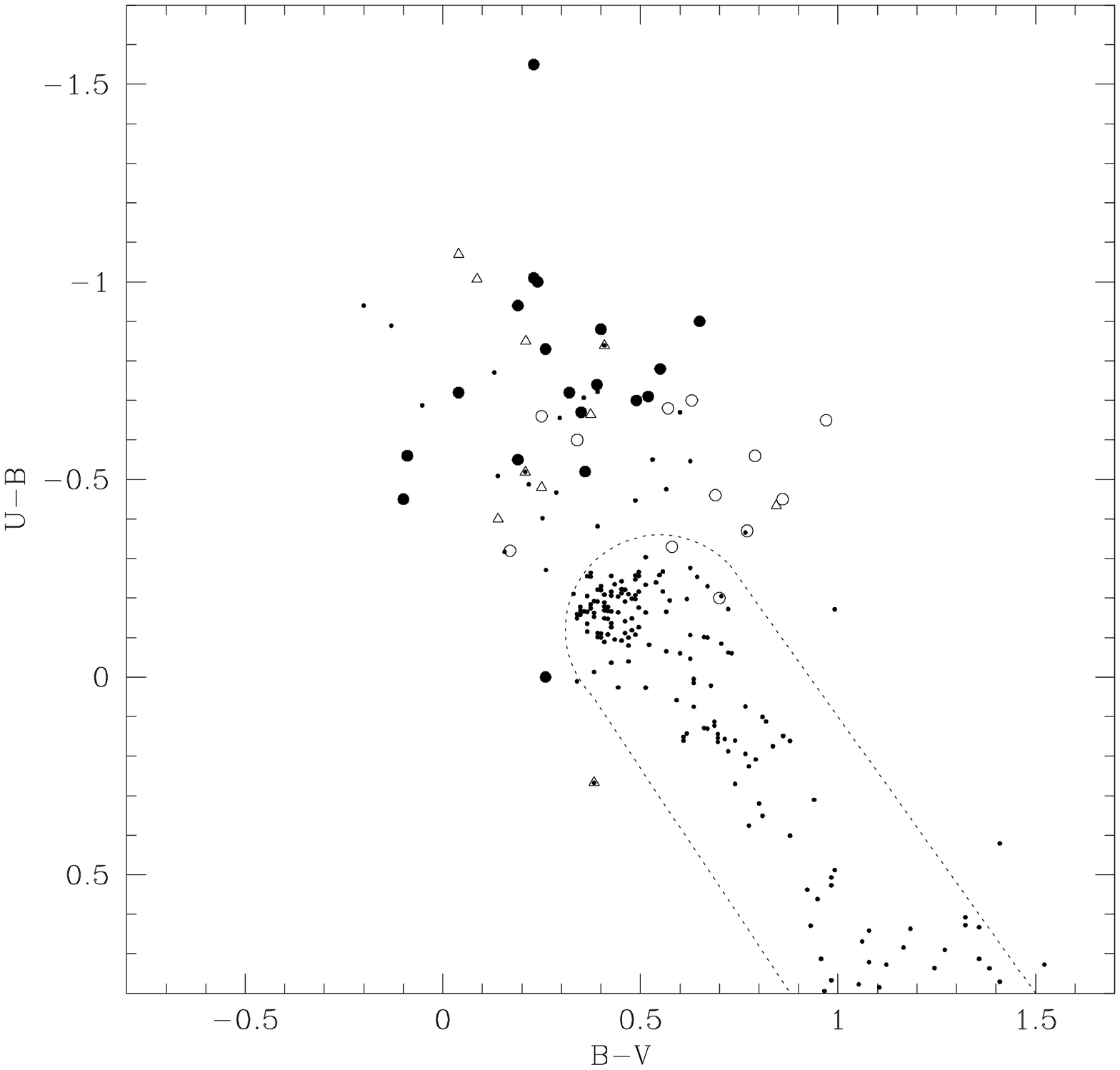}
\caption[]{
U-B versus B-V for all the AGNs in our field.  
The different symbols represent AGNs detected in X-ray and already present 
in the optical sample (solid circles), AGNs detected in X-ray and not 
present in the optical sample (empty circles), AGNs not detected in X-ray 
(triangles). The small dots represent all the point--like objects in our
CCD data with $m_B \le$ 22.5. The dotted curve shows the approximate locus
occupied by stars in this plane. 
}
\end{figure}

Having colours for all the X-ray selected AGNs, we can also estimate how many of 
them would have been missed by a purely optical selection. Figure 8
shows, with different symbols (see figure caption) all known AGNs in our field. 
The small dots represent all the point--like objects in our
CCD data with $m_B \le$ 22.5. The dotted curve shows the approximate locus
occupied by stars in this plane. This figure shows that
two X-ray selected AGNs (X251--37 at z = 2.71 and X042--11 at z = 1.062) are 
inside the star locus and therefore would have not been easily selected as AGN 
candidates on the basis of this diagram.
Moreover, two X-ray selected AGNs (X233--45 at z = 1.180, and X409--28, 
the broad line radio galaxy at z = 0.958) are classified as extended by our 
morphological algorithm and two more (X043--12, the AGN2 at z = 2.80, and 
X211--22 at z = 0.281) have uncertain classification (i.e. p/e in Table 3).
Since the colour-colour area occupied by faint AGNs contains a number of
extended objects which is about ten times higher than that of the point--like
ones, also these four objects would have not been easily selected as AGN
candidates by purely optical data. Moreover, the location of these objects in
the magnitude -- redshift plane is not the same as that of AGNs which are more
easily selected optically; five have $m_B \ge$ 22.2 and only one has
$m_B <$ 22.2, while the corresponding numbers for all the other AGNs with 
redshift in this field are 11 and 25. The same difference is also visible
as a function of the absolute magnitude: five of the objects which would
have been more difficult to detect from the optical data have $M_B \ge$ -22.4 
and only one has $M_B <$ -22.4, while the corresponding numbers for all the 
other AGNs are 14 and 22. This clearly shows that not only colour--colour 
selection of AGN candidates among point--like objects can be significantly 
incomplete at faint apparent magnitudes, but also the incompleteness increases 
at faint absolute magnitudes. As a consequence, if these not so stellar AGNs
were really missed in faint optical surveys, this would introduce a bias
in the derived optical luminosity function, which would appear flatter
than the real one. Finally, we note also that a not negligible number of 
X-ray selected AGNs are significantly redder in B-V than the bulk of AGNs
(see figure 8). Also this has to be taken into account in devising 
the most efficient optical selection criteria for faint AGN candidates.

\section{ Conclusion}

We have presented the X-ray data and the optical identifications for
a deep ROSAT PSPC observation in the Marano field. Careful statistical
analysis of multi--colour CCD data in the error boxes of the 50 X-ray sources 
detected in the inner region of the ROSAT field (15\arcm \ radius) has
led to the identification of 42 sources, corresponding to 84\% of the
X-ray sample. These 42 reliable identifications are 33 AGNs (including 
two radio galaxies and one BL Lac candidate; 79\% of the identified sources), 
2 galaxies, 3 groups or clusters of galaxies and 4 stars. Except for the 
higher fraction of unidentified sources, the identification content of this 
sample is in excellent agreement with what has been found, at a similar flux
limit, in the much deeper PSPC and HRI surveys in the Lockman field (Hasinger
et al. 1998, Schmidt et al. 1998).

With simple simulations we have shown that in a few cases the reason for
not having found an optical identification can be due to the fact that a 
single X-ray source in our list may be produced by 
two different close--by sources, so that its detected position may be 
significantly wrong. Most of the unidentified sources have a large ratio
of X-ray to optical fluxes and harder than average X-ray spectra. Since
most of the identified objects with these characteristics in our field and
in the Lockman field are AGNs, we conclude that most of the sources with 
good position determination but without identification are likely to be AGNs.

Most of the sources in the Marano and Lockman fields occupy well defined
bands in the plane hardness ratio versus X-ray flux. These bands correspond
to the same energy spectral index range 0.5 $\le \alpha_x \le $ 2.0,
only if the effective X-ray absorbing column in the Lockman field is
about twice the value of atomic hydrogen derived from measurements in the 
21 cm line. In both samples there are about (15--20)\% of the sources which 
appear to have hard or absorbed spectra, with HR values close to 1 and,
especially in the Marano field, they tend to be more numerous at low flux
levels. However, since there is an obvious selection effect which favours hard
spectra near the detection limit, it is difficult to quantify the reality
of this effect. The fraction of sources without optical 
identification is significantly higher among these hard sources than among 
sources with ``normal'' X-ray spectrum. The eleven identifications (Lockman and
Marano fields together) among these hard sources (9 AGNs and 2 galaxies)
show that AGNs are the dominant population also in this range of HR. 
However, contrary to what happens for objects with ``normal'' X-ray spectrum, 
a relatively large fraction of them does not show the prominent broad lines
typical of classical quasars. 

Finally, comparing the optically and X-ray selected samples of AGNs in the
same area, we estimate that the ``efficiency'' of AGN selection with X-ray
exposures reaching about $4 \times 10^{-15}$ \erg is
$\sim$ 65\% and $\sim$ 20\% in the magnitude ranges $m_B < 22.5$ and 
$22.5 < m_B < 23.5$, respectively. On the other hand, a not negligible
fraction of the X-ray selected AGNs would have not been easily selected
as AGN candidates on the basis of purely optical criteria, either because
of colours similar to those of normal stars or because of morphological
classification not consistent with point--like sources.

Moreover, the location of these objects in the
magnitude -- redshift plane is not the same as that of AGNs which are more
easily selected optically. They tend to be fainter in terms of both apparent
and absolute magnitudes. As a consequence, if these not so stellar AGNs
were really missed in faint optical surveys, this would introduce a bias
in the derived optical luminosity function, which would appear flatter
than the real one. Finally, we note also that a not negligible number of 
X-ray selected AGNs are significantly redder in B-V than the bulk of AGNS.
Also this has to be taken into account in devising 
the most efficient optical selection criteria for faint AGN candidates.
\begin{acknowledgements}
The ROSAT project is supported by the Bunderministerium f\"ur Forschung und
Technologie (BMFT), by the Science and Engineering Research Council (SERC)
and by the National Aeronautics and Space Administration (NASA). This work
was supported in part by NASA grants NAG5-1531 (M.S.), NAG8-794, NAG5-1649 and
NAGW-2508 (R.B. and R.G). G.H. acknowledges the DARA grant FKZ 50 OR 9403 5;
G.Z. acknowledges partial support by the Italian Space Agency (ASI) under
ASI contract ARS-96-70 and by the Italian Ministry for University and Research
(MURST) under grant Cofin98-02-32.
\end{acknowledgements}
%
%

%
\end{document}